\documentclass[aps,prd,a4paper,twocolumn,amsmath,showpacs,superscriptaddress,nofootinbib,preprintnumbers]{revtex4-1}
\pdfoutput=1

\usepackage{amssymb,amsmath,latexsym,mathrsfs}
\usepackage[sort&compress]{natbib}
\usepackage{graphicx,subfigure}
\usepackage{epsfig}
\usepackage{varioref,xr-hyper}
\usepackage{color}
\usepackage{multirow}
\usepackage{array}
\usepackage{hyperref}
\usepackage{wasysym}
\usepackage{color}
\usepackage{float}
\usepackage{xcolor}
\usepackage[utf8]{inputenc}
\usepackage[T1]{fontenc}

\begin{document}

\title{Dawn of the dark: unified dark sectors and the EDGES Cosmic Dawn 21-cm signal}

\author{Weiqiang Yang}
\email{d11102004@163.com}
\affiliation{Department of Physics, Liaoning Normal University, Dalian, 116029, P. R. China}

\author{Supriya Pan}
\email{supriya.maths@presiuniv.ac.in}
\affiliation{Department of Mathematics, Presidency University, 86/1 College Street, Kolkata 700073, India}

\author{Sunny Vagnozzi}
\email{sunny.vagnozzi@ast.cam.ac.uk}
\affiliation{Kavli Institute for Cosmology (KICC) and Institute of Astronomy, University of Cambridge, Madingley Road, Cambridge CB3 0HA, United Kingdom}

\author{Eleonora Di Valentino}
\email{eleonora.divalentino@manchester.ac.uk}
\affiliation{Jodrell Bank Center for Astrophysics, School of Physics and Astronomy, University of Manchester, Oxford Road, Manchester M13 9PL, United Kingdom}

\author{David F. Mota}
\email{d.f.mota@astro.uio.no}
\affiliation{Institute of Theoretical Astrophysics, University of Oslo, P.O. Box 1029 Blindern, N-0315 Oslo, Norway}

\author{Salvatore Capozziello}
\email{capozziello@na.infn.it}
\affiliation{Dipartimento di Fisica, Universit\`{a} di Napoli Federico II, Complesso Universitario di Monte Sant'Angelo, Via Cinthia, 21, I-80126 Napoli, Italy}
\affiliation{Istituto Nazionale di Fisica Nucleare (INFN), Sezione di Napoli, Complesso Universitario di Monte Sant'Angelo, Via Cinthia, 21, I-80126 Napoli, Italy}
\affiliation{Gran Sasso Science Institute (GSSI), Viale Francesco Crispi 7, I-67100 L’Aquila, Italy}
\affiliation{Tomsk State Pedagogical University, ul. Kievskaya, 60, 634061 Tomsk, Russia}
%---------------------------------------------------------
%--------------------------------------------------------
\begin{abstract}
While the origin and composition of dark matter and dark energy remains unknown, it is possible that they might represent two manifestations of a single entity, as occurring in unified dark sector models. On the other hand, advances in our understanding of the dark sector of the Universe might arise from Cosmic Dawn, the epoch when the first stars formed. In particular, the first detection of the global 21-cm absorption signal at Cosmic Dawn from the EDGES experiment opens up a new arena wherein to test models of dark matter and dark energy. Here, we consider generalized and modified Chaplygin gas models as candidate unified dark sector models. We first constrain these models against Cosmic Microwave Background data from the \textit{Planck} satellite, before exploring how the inclusion of the global 21-cm signal measured by EDGES can improve limits on the model parameters, finding that the uncertainties on the parameters of the Chaplygin gas models can be reduced by a factor between $1.5$ and $10$. We also find that within the generalized Chaplygin gas model, the tension between the CMB and local determinations of the Hubble constant $H_0$ is reduced from $\approx 4\sigma$ to $\approx 1.3\sigma$. In conclusion, we find that the global 21-cm signal at Cosmic Dawn can provide an extraordinary window onto the physics of unified dark sectors.
\end{abstract}

\date{\today}
\maketitle

\section{Introduction}
\label{sec:introduction}

A combination of cosmological observations have depicted a rather odd picture of our Universe, whose energy content resides mostly in two dark components: a cold clustering dust-like component responsible for the formation of structure, usually referred to as dark matter (DM), and a smooth component whose properties very much resemble that of vacuum energy, responsible for the inferred late-time acceleration of the Universe and usually referred to as dark energy (DE)~\cite{Riess:1998cb,Perlmutter:1998np,Ester,Alam:2016hwk,Troxel:2017xyo,Aghanim:2018eyx}. Unraveling the nature and origin of DM and DE is perhaps one of the major challenges physics has ever faced.

The literature abounds with theoretical and phenomenological models for the dark sector. Most DM models rely on the introduction of additional particles and/or forces usually weakly coupled to the Standard Model (see for instance~\cite{Dodelson:1993je,Jungman:1995df,Cirelli:2005uq,McDonald:2007ex,ArkaniHamed:2008qn,Visinelli:2009zm,Feng:2009mn,Kaplan:2009de,Fan:2013yva,Petraki:2014uza,Foot:2014uba,Foot:2014osa,Foot:2016wvj,Visinelli:2017imh,Visinelli:2017qga,Tenkanen:2019xzn}), although models exist where the dynamics usually attributed to DM emerge from modifications to Einstein's theory of General Relativity (see e.g.~\cite{Milgrom:1983ca,Capozziello:2006ph,Boehmer:2007kx,Chamseddine:2013kea,Myrzakulov:2015nqa,Myrzakulov:2015kda,Luongo,Rinaldi:2016oqp,Verlinde:2016toy,Capozziello:2017rvz,Vagnozzi:2017ilo,Sergey,deHaro:2018sqw,Khalifeh:2019zfi}). On the other hand, several DE models rely on the introduction of a new fluid, possibly through a very light field (see for example~\cite{Ratra:1987rm,Caldwell:1997ii,Zlatev:1998tr,Freese:2002sq,Cai:2009zp,Hlozek:2014lca,Cognola:2016gjy,Vagnozzi:2018jhn,Casalino:2018tcd,Visinelli:2018utg,DiValentino:2019exe,Visinelli:2019qqu}), or through modifications to GR (see e.g.~\cite{Li:2004rb,Nojiri:2006ri,Capozziello:2006uv, Viability, Report, Felix, Hu:2007nk,Starobinsky:2007hu,Appleby:2007vb,Cognola:2007zu,Jhingan:2008ym,Saridakis:2009bv,Appleby:2009uf,Dent:2011zz,Myrzakulov:2015qaa,Cai:2015emx,Sebastiani:2016ras,Dutta:2017fjw,Casalino:2018wnc}). Although within most dark sector models the two dark components do not interact, an interesting class of cosmological models (usually referred to as coupled DE) feature interactions between DM and DE (see for example~\cite{Amendola:1999er,Barrow:2006hia,He:2008tn,Valiviita:2008iv,Gavela:2009cy,Martinelli:2010rt,Pan:2012ki,Yang:2014gza,yang:2014vza,Yang:2014hea,Tamanini:2015iia,Murgia:2016ccp,Nunes:2016dlj,Yang:2016evp,Pan:2016ngu,Shafieloo:2016bpk,Sharov:2017iue,Kumar:2017dnp,DiValentino:2017iww,Yang:2017yme,Yang:2017ccc,Yang:2017zjs,Pan:2017ent,Yang:2018euj,Yang:2018ubt,Yang:2018pej,Yang:2018xlt,Yang:2018uae,Yang:2018qec,Paliathanasis:2019hbi,Pan:2019jqh,Li:2019san,Yang:2019bpr,Yang:2019vni,Yang:2019uzo,DiValentino:2019ffd,Benetti:2019lxu,Mukhopadhyay:2019jla,Carneiro:2019rly,Kase:2019veo,Yang:2019uog,DiValentino:2019jae}). While we cannot do justice to the enormous literature on dark sector models, it is clear from the discussion so far that there is no shortage of proposed models for DM and DE and observational tests thereof. For a recent discussion on this point, see \cite{Rocco}.

Despite decades of experimental efforts in detecting DM and DE, the two dark actors of the play which is the evolution of our Universe have so far eluded our understanding. Therefore, it is perhaps timely to consider alternatives beyond the most studied models, and explore whether they can be tested with current observations. In this sense, there exists another interesting possibility as dark sector models go: it is possible that DM and DE might be two manifestations of the same underlying entity, such as a single fluid whose behaviour first mimics that of DM and at late times that of DE (see e.g. the case of the Anton-Schmidt fluid recently adopted in cosmology \cite{Anton1,Anton2}).

It is with a particular class of such unified dark sector models, known as Chaplygin gases, that we shall be concerned in this work. Chaplygin gases are models featuring an exotic equation of state (EoS) relating pressure $p$ and energy density $\rho$ of a fluid. The Chaplygin gas (CG) was first introduced in a cosmological context in~\cite{Kamenshchik:2001cp}, although its origin date back to 1904 when russian scientist Sergey Chaplygin first studied it in the context of aerodynamics~\cite{Chaplygin}. In the original CG model~\cite{Kamenshchik:2001cp}, the CG EoS is given by $p =- A/\rho$, with $A>0$ an arbitrary constant. It was soon realized that the CG model is interesting from both the theoretical and cosmological points of view. In fact, the CG shares a close connection to string theory, as it emerges from the Nambu-Goto action for $d$-branes moving in a $(d+2)$-dimensional spacetime in lightcone parametrization~\cite{Bordemann:1993ep}, while it is the only known fluid admitting a supersymmetric generalization~\cite{Jackiw:2000cc}.

Following the seminal work of~\cite{Kamenshchik:2001cp}, CG models were further studied and extended. In this regard, two important extensions known as generalized Chaplygin gas and modified Chaplygin gas were presented in~\cite{Bento:2002ps} and~\cite{Benaoum:2002zs} respectively. Since these two models will be central to our work, we shall discuss them in more detail in Sec.~\ref{sec:unified}. As realized in~\cite{Kamenshchik:2001cp}, the CG effectively interpolates between a dust-like (and hence DM-like) behaviour at high redshift ($\rho \propto a^{-3}$, with $a$ the scale factor of the Universe), and a cosmological constant-like behaviour at low redshift ($\rho \propto {\rm const}$), and hence constitutes a potentially interesting unified model for the dark sector of the Universe. These features are also shared by the generalized and modified Chaplygin gas models. Moreover, in between the DM-like and DE-like epochs, CG models usually feature an exotic epoch of soft or stiff matter domination (\textit{i.e.} with equation of state close to $1$) which is otherwise absent in $\Lambda$CDM. Over the years, several works have examined theoretical, phenomenological, and observational aspects of CG models and their extensions. For an incomplete list of works, see e.g.~\cite{Bilic:2001cg,Dev:2002qa,Gorini:2002kf,Makler:2002jv,Bento:2002yx,Alcaniz:2002yt,Bento:2003we,Amendola:2003bz,Dev:2003cx,Bertolami:2004ic,Debnath:2004cd,Zhang:2004gc,Sen:2005sk,Zhang:2005jj,Wu:2006pe,BouhmadiLopez:2007ts,Gorini:2007ta,Ali:2010sv,Xu:2010zzb,Lu:2010zzf,Xu:2012qx,Xu:2012ca,Campos:2012ez,Wang:2013qy,Khurshudyan:2014ewa,Avelino:2014nva,Sharov:2015ifa,Khurshudyan:2015mva,vonMarttens:2017njo,Yang:2019jwn}. Observational studies have shown that Chaplygin gas models are in agreement with data and hence can constitute a viable candidate for the dark sector.

A word of caution concerning Chaplygin gas models is required at this point. Early works argued that Chaplygin gases, and in particular the adiabatic version thereof, are unstable at the perturbative level, with strong oscillations appearing in the matter power spectrum~\cite{Sandvik:2002jz} and the CMB anisotropy spectrum~\cite{Amendola:2003bz}. Nonetheless, various solutions have been considered in the literature, ranging from entropic~\cite{Reis:2003mw} or non-adiabatic perturbations~\cite{HipolitoRicaldi:2009je}, to decompositions into interacting DM and DE-like terms~\cite{Bento:2004uh,HipolitoRicaldi:2010mf,Wang:2013qy}, and combinations of the latter solutions~\cite{Zimdahl:2005ir,Borges:2013bya,Carneiro:2014jza}. However, another possibility pursued in the literature~\cite{Xu:2012qx,Xu:2012ca} has been that of still considering pure adiabatic perturbations, but restricting the parameter space of the models, at the data analysis stage, by imposing the viability condition that the squared sound speed of the dark fluid be strictly positive. As argued in~\cite{Xu:2012qx,Xu:2012ca}, this is another viable way of ensuring that the Chaplygin gas models are perturbatively stable. This is the approach we will be following in the rest of the paper.

One particularly interesting cosmological probe, which has the potential to revolutionize our understanding of the Universe, is the 21-cm line of neutral Hydrogen, related to the hyperfine splitting of the Hydrogen atom 1s ground state and caused by the interaction between the electron and proton magnetic moments. Within the redshift range $15 \lesssim z \lesssim 25$ known as Cosmic Dawn, during which the first stars formed, the UV photons thereby emitted excited this hyperfine transition, sourcing 21-cm absorption against the Cosmic Microwave Background (CMB)~\cite{Wouthuysen,Field,Hirata:2005mz}. Detecting this global 21-cm absorption signal was the main goal of the Experiment to Detect the Global EoR Signature (EDGES), which succeeded in 2018~\cite{Bowman:2018yin}.

The EDGES detection, albeit revealing a signal twice as large as standard expectations, provides a unique window into the high-redshift Universe ($z \approx 17$), otherwise inaccessible to more traditional tracers of the large-scale structure. Therefore, EDGES might be uniquely placed to probe dark sector components which are either non-standard or exhibit non-standard behaviour at high redshift, while being completely consistent with CMB and low-redshift measurements. Chaplygin gases are particularly intriguing in this sense: their non-standard behaviour between the DM-like and DE-like epochs might show up at Cosmic Dawn and consequently in the global 21-cm absorption signal. Conversely, the global 21-cm absorption signal might provide novel and valuable constraints on CG models, complementary to the more traditional CMB and large-scale structure probes. Our goal in this work is to reassess the observational status of CG models in light of the global 21-cm absorption signal detected by EDGES. Focusing on generalized and modified Chaplygin gas models, we will use this signal to provide new constraints on the parameters of these models.

The rest of this paper is then organized as follows. In Sec.~\ref{sec:unified}, we discuss the cosmology of unified dark sector models, focusing on the generalized and modified Chaplygin gas models. In particular, we discuss in detail the background and perturbation evolution therein. In Sec.~\ref{sec:21cm} we describe in more detail the global 21-cm absorption signal at Cosmic Dawn, as well as its first detection by EDGES. We then proceed to discuss the data and analysis methods we use in Sec.~\ref{sec:data}, before discussing our results in Sec.~\ref{sec:results} and finally drawing concluding remarks in Sec.~\ref{sec:conclusions}.

\section{Cosmology of unified dark sectors}
\label{sec:unified}

In the following, we shall describe in more detail the two unified dark sector models we will consider in this work. We begin by describing in more detail the two models and their background evolution in Sec.~\ref{subsec:background}. In particular, we discuss the generalized Chaplygin gas in Sec.~\ref{subsubsec:gcg} and the modified Chaplygin gas in Sec.~\ref{subsubsec:mcg}. We then go on to discuss the evolution of perturbations in Sec.~\ref{subsec:perturbations}.

\subsection{Background evolution}
\label{subsec:background}

We work within the framework of a homogeneous and isotropic Universe, whose geometry is described  by a spatially flat Friedmann$-$Lema\^{i}tre$-$Robertson$-$Walker (FLRW) line element, characterized by the scale factor as a function of conformal time $a(\tau)$. We assume that the gravitational sector, to which the matter sector is minimally coupled, is described by Einstein's General Relativity. We further assume that the energy budget of the Universe comes in the form of four species: baryons, photons, neutrinos, and a unified dark fluid (UDF). The UDF will behave as dark matter (DM), dark energy (DE), or a different type of fluid as the Universe expands~\footnote{In the following, we shall fix the total neutrino mass to $M_{\nu}=0.06\,{\rm eV}$, the minimal value allowed within the normal ordering as done in the \textit{Planck} baseline analyses. This is justified by the current very tight upper limits on $M_{\nu}$~\cite{Palanque-Delabrouille:2015pga,Giusarma:2016phn,Vagnozzi:2017ovm,Giusarma:2018jei,Aghanim:2018eyx,Vagnozzi:2019utt}, which also mildly favour the normal ordering~\cite{Vagnozzi:2017ovm,Simpson:2017qvj,Schwetz:2017fey}.}. Let us denote the energy densities and pressures of these species by $\rho_i$ and $p_i$ respectively, where $i=b,\gamma,u$ for baryons, photons, and the UDF respectively. We assume that the fluids do not share interactions between each other aside from the gravitational ones. In other words, each fluid $i$ obeys a separate continuity equation (which follows from the Bianchi identities) given by:
\begin{eqnarray}
\dot{\rho}_i + 3 \frac{\dot{a}}{a} (p_i + \rho_i) =  0\,.
\label{eq:continuity}
\end{eqnarray}
For baryons ($p_b = 0$) and photons ($p_\gamma = \rho_\gamma/3$), Eq.~(\ref{eq:continuity}) is trivially solved to give $\rho_b \propto a^{-3}$ and $\rho_{\gamma} \propto a^{-4}$ respectively. For the UDF, the solution to Eq.~(\ref{eq:continuity}) depends on the specific form of the equation of state, which specifies $p_u$ as a function of $\rho_u$: $p_u = f(\rho_u)$.  

To make progress we need to specify the functional form of the equation of state of the UDF. As anticipated in Sec.~\ref{sec:introduction}, in this work we shall consider two particular UDF models which go under the names of generalized Chaplygin gas (GCG) and modified Chaplygin gas (MCG), and which will be described in Sec.~\ref{subsubsec:gcg} and Sec.~\ref{subsubsec:mcg} respectively.

\subsubsection{Generalized Chaplygin gas}
\label{subsubsec:gcg}

We begin by considering the generalized Chaplygin gas (GCG) model for an unified dark fluid. The GCG model is characterized by the following equation of state relating its pressure $p_{\rm gcg}$ and its energy density $\rho_{\rm gcg}$:
\begin{eqnarray}
p_{\rm gcg} = -\frac{A}{(\rho_{\rm gcg})^{\alpha}}\,,
\label{eq:gcg}
\end{eqnarray}
where $A$ and $\alpha$ are two real constants. The original Chaplygin gas model discussed in Sec.~\ref{sec:introduction} is recovered when $\alpha= 1$. We can then solve the continuity equation Eq.~(\ref{eq:continuity}) to find the evolution of the GCG energy density as a function of the scale factor, given by the following:
\begin{eqnarray}
\rho_{\rm gcg}(a)=\rho_{{\rm gcg},0} \left ( B_{s}+(1-B_{s})a^{-3(1+\alpha)} \right )\,
^{\frac{1}{1+\alpha}},
\label{eq:rhogcg}
\end{eqnarray}
where $\rho_{{\rm gcg},0}$ denotes the energy density of the GCG fluid today, and $B_{s}=A\rho_{{\rm gcg},0}^{-(1+\alpha)}$. The evolution of the GCG equation of state $w_{\rm gcg}$ as a function of the scale factor can easily be found by combining Eqs.~(\ref{eq:gcg}, \ref{eq:rhogcg}) to give the following:
\begin{eqnarray}
w_{\rm gcg}(a)=-\frac{B_{s}}{B_{s}+(1-B_{s})a^{-3(1+\alpha)}}\,.
\label{eq:eosgcg}
\end{eqnarray}
Eq.~(\ref{eq:rhogcg}) fully specifies the background evolution in the presence of a GCG fluid (together with the usual baryon, photon and neutrino components).

It is easy to see, as was first shown in~\cite{Bento:2002ps}, that the GCG fluid interpolates between a dust-dominated universe ($\rho_{\rm gcg} \propto a^{-3}$) and a de Sitter phase ($\rho_{\rm gcg} \propto {\rm const}$), via an intermediate epoch of soft matter domination (wherein $p \approx \alpha \rho$, with this intermediate epoch being one of stiff matter domination when $\alpha=1$ and the original Chaplygin gas model is recovered): it is this feature of interpolating between an effective DM component and an effective DE one which makes the GCG an appealing model for an unified dark sector.

\subsubsection{Modified Chaplygin gas}
\label{subsubsec:mcg}

We then consider the modified Chaplygin gas (MCG) model for an unified dark fluid. The relation between pressure $p_{\rm mcg}$ and energy density $\rho_{\rm mcg}$ for the MCG model is given by the following: 
\begin{eqnarray}
p_{\rm mcg} =  B \rho_{\rm mcg} - \frac{A}{(\rho_{\rm mcg})^{\alpha}}\,,
\label{eq:mcg}
\end{eqnarray}
where again $A$, $B$, and $\alpha $ are three real constants. When setting $A=0$, the MCG behaves as a perfect fluid with equation of state $w=B$, whereas setting $B=0$ one recovers the GCG model, and further setting $\alpha=0$ the original Chaplygin gas model. Solving the continuity equation Eq.~(\ref{eq:continuity}) we find that the evolution of the MCG energy density is given by the following:
\begin{eqnarray}
\hskip -0.5 cm \rho_{\rm mcg}(a)=\rho _{{\rm mcg},0} \left ( B_{s}+(1-B_{s})a^{-3(1+B)(1+\alpha)} \right )^{\frac{1}{1+\alpha}}\,,
\label{eq:rhomcg}
\end{eqnarray}
where $\rho_{{\rm mcg},0}$ denotes the energy density of the MCG fluid today, and $B_{s}=A\rho_{{\rm gcg},0}^{-(1+\alpha)}/(1+B)$. As we did for the GCG model, we can combine Eqs.~(\ref{eq:mcg},\ref{eq:rhomcg}) to find the evolution of the MCG equation of state $w_{\rm mcg}$ as a function of the scale factor, which is given by the following:
\begin{eqnarray}
w_{\rm mcg}(a)=B-\frac{B_{s}(1+B)}{B_{s}+(1-B_{s})a^{-3(1+B)(1+\alpha)}}\,.
\label{eq:eosmcg}
\end{eqnarray}
Eq.~(\ref{eq:rhomcg}) fully specifies the background evolution in the presence of a MCG fluid (together with the usual baryon, photon and neutrino components).

Being a generalization of the GCG model, also the MCG model interpolates between a dust-dominated Universe (or more generally an Universe dominated by a perfect fluid with chosen EoS) and a de Sitter phase, via an intermediate epoch of soft matter domination.

\subsection{Evolution of perturbations}
\label{subsec:perturbations}

Once the parameters defining the GCG and MCG models are chosen, the background evolution is fully specified by Eqs.~(\ref{eq:rhogcg},\ref{eq:rhomcg}), or equivalently Eqs.~(\ref{eq:eosgcg},\ref{eq:eosmcg}). Knowledge of the background evolution is in principle sufficient to predict the theoretical global 21-cm absorption signal given a set of GCG or MCG parameter, prediction which we can then compare against data, in this case the measurement from EDGES. However in this work we intend to constrain the GCG and MCG models not only in light of the EDGES measurement, but also using measurements of temperature and polarization anisotropies in the CMB. The reason is that CMB measurements are extremely efficient in breaking a number of parameter degeneracies which other probes alone (including EDGES) would not be able to. To be able to predict the theoretical values for the CMB temperature and polarization anisotropy spectra given a set of GCG or MCG parameters requires us to be able to track the evolution of perturbations in the GCG and MCG fluids. Therefore, in this section we describe the evolution of perturbations within the two models we have considered.

To make progress, we work within the synchronous gauge, wherein the perturbed FLRW line element takes the form~\cite{Ma:1995ey}:
\begin{eqnarray}
ds^2 = a^2(\tau) \left [- d\tau^2 + \left ( \delta_{ij} + h_{ij} \right ) dx^i d^j \right ]\,,
\label{eq:synchronous}
\end{eqnarray}
where $\tau$ denotes conformal time and $h_{ij}$ denotes the synchronous gauge metric perturbation. Within this gauge and neglecting shear stress, consistently with the earlier work of~\cite{Xu:2012zm}, we can track the evolution of the Fourier space UDF density perturbation $\delta_u$ and velocity divergence $\theta_u$. In the usual notation of Ma \& Bertschinger~\cite{Ma:1995ey}, the evolution equations for $\delta_u$ and $\theta_u$ are given by:
\begin{eqnarray}
\label{eq:deltau}
\dot{\delta}_u &=& -(1+w_u) \left(\theta_u + \frac{\dot{h}}{2} \right )- 3 \mathcal{H} \left(c_s^2 - w_u\right) \delta_u, \\
\label{eq:thetau}
\dot{\theta}_u &=& -\mathcal{H}\left(1- 3 c^2_{s}\right)\theta_u + \frac{c^2_{s}}{1+w_u} k^2 \delta_u\,,
\end{eqnarray}
where the overdot denotes differentiation with respect to conformal time, the conformal Hubble rate is given by ${\cal H} = \dot{a}/a$, $h \equiv h^{j}_{j}$ is the trace of the metric perturbation $h_{ij}$, $w_u$ denotes the effective EoS of the unified dark fluid [given by Eq.~(\ref{eq:eosgcg}) or Eq.~(\ref{eq:eosmcg}) depending on whether one is considering the GCG or MCG model], and $c_s^2$ is the squared sound speed of the unified dark fluid.

Following~\cite{Xu:2012qx,Xu:2012ca}, we consider pure adiabatic contributions to the perturbations. As we discussed in Sec.~\ref{sec:introduction}, this is potentially problematic due to instabilities, but we will follow the approach of~\cite{Xu:2012qx,Xu:2012ca} to ensure that the models we consider are perturbatively stable. Under these assumptions, the squared sound speed for the GCG fluid is given by:
\begin{eqnarray}
c_s^2 = \delta p_u/\delta \rho_{u} = -\alpha w_{\rm gcg}\,.
\label{eq:cs2gcg}
\end{eqnarray}
Similarly, the squared sound speed for the MCG fluid is given by:
\begin{eqnarray}
c_s^2  = \delta p_u/\delta \rho_{u} = - \alpha w_{\rm mcg} + (1+\alpha) B\,.
\label{eq:cs2mcg}
\end{eqnarray}
In order for perturbations in the UDF to be stable, one must ensure that $c_s^2 > 0$. In the case of the GCG, we follow~\cite{Xu:2012qx} and require $\alpha > 0$ and $0 < B_s < 1$. In such a way, $c_s^2$ as in Eq.~(\ref{eq:cs2gcg}) is strictly positive and perturbations in the GCG are stable. For the MCG case, we follow~\cite{Xu:2012ca} and ensure that $c_s^2 > 0$ in our Markov Chain Monte Carlo analysis (to be discussed in more detail in Sec.~\ref{sec:data}) by rejecting points with $c_s^2<0$, where $c_s^2$ is given by Eq.~(\ref{eq:cs2mcg}). Effectively, this corresponds to imposing the additional constraint $B > -\alpha B_s/(1+\alpha B_s)$ at the level of prior. In such a way, perturbations in the MCG are stable.

\section{EDGES and the global 21-cm absorption signal at Cosmic Dawn}
\label{sec:21cm}

We now briefly review the physics underlying the global 21-cm signal at Cosmic Dawn, before discussing the detection of such signal by the EDGES experiment. We begin by discussing the theory behind the signal in Sec.~\ref{subsec:theory}, before discussing the EDGES detection in Sec.~\ref{subsec:edges}.

\subsection{Theory}
\label{subsec:theory}

The 21-cm signal is a unique probe of the evolution of neutral Hydrogen, which in turn is an extremely useful tracer of the properties of the gas across cosmic time. In particular, it is an extraordinary probe of the so-called Cosmic Dawn, the period when the very first sources of light formed and ended the Dark Ages (the era which began with the formation of the CMB). We encourage the interested reader to consult more detailed reviews on the subject such as~\cite{Furlanetto:2006jb,Pritchard:2011xb,Barkana:2016nyr}.

The 21-cm line responsible for this signal is produced by the hyperfine splitting of the 1s ground state of the Hydrogen atom, caused by the interaction between the electron and proton magnetic moments. These interactions lead to a hyperfine splitting between the so-called singlet and triplet states, whose energy levels are separated by $\Delta E = 5.9 \times 10^{-6}\,{\rm eV}$ (which corresponds to a rest wavelength of $\approx 21\,{\rm cm}$).

The relative abundance of singlet and triplet states is captured by the so-called spin temperature $T_s$. While $T_s$ is not, strictly speaking, a true thermodynamic temperature, it nonetheless provides very useful insight into the 21-cm signal. Working within the Rayleigh-Jeans limit (relevant for the frequencies in question, which are far from the peak frequency of the CMB), the strength of the 21-cm signal is typically quantified through a so-called differential brightness temperature (relative to the CMB) $T_{21}$, whose evolution with redshift $z$ is given by~\cite{Zaldarriaga:2003du}:
\begin{eqnarray}
\label{21cm}
\hskip -0.55 cm T_{21} (z) \approx \left ( \frac{2c^3 \hbar A_{10} n_{\rm HI}}{16 k_B \nu^2_{0}} \right)  \frac{1}{H(z)} \left (\frac{1 - T_{\gamma}(z)/T_s(z)}{1+z} \right)\,,
\label{eq:21}
\end{eqnarray}
with $T_{\gamma}(z)$ the temperature of CMB photons as a function of redshift, $c$ the speed of light, $\hbar$ the reduced Planck constant, $k_B$ Boltzman's constant, $A_{10} =2.85 \times 10^{-15}\,{\rm s}^{-1}$ the emission coefficient for the hyperfine triplet-single transition whose rest-frame transition frequency is $\nu_{0} = 1420.4\,{\rm MHz}$, $n_{\rm HI}$ the number density of neutral Hydrogen (well approximated by $n_{\rm HI} \simeq 1-x_e$, where $x_e$ is the ionization fraction). If at a certain redshift $T_s<T_{\gamma}$, then $T_{21}<0$ and the 21-cm signal is seen in absorption. Conversely, at redshifts for which $T_s>T_{\gamma}$, $T_{21}>0$ and the 21-cm signal will be seen in emission. 

The evolution of the 21-cm signal with cosmic time is governed by the interplay between three temperatures: aside from the already introduced $T_s$ and $T_{\gamma}$, the third relevant temperature is the gas kinetic temperature $T_k$. At early times ($z \gtrsim 200$), although the fraction of free electrons is very low, it is still high enough to thermally couple the gas to the CMB photons, whereas the high gas density makes collisional coupling very effective. These two factors combine to ensure that $T_s=T_k=T_{\gamma}$, leading to no detectable 21-cm signal. However, at $z \approx 200$ the gas decouples from the CMB and starts cooling adiabatically. For a while collisional coupling remains effective, so that $T_s \approx T_k < T_{\gamma}$, leading to $T_{21}<0$ and hence a very early 21-cm absorption signal. However, at $z \approx 40$ the gas density has become too low for collisional coupling to be effective, and radiation coupling couples the spin temperature to the CMB, so that $T_k<T_s \approx T_{\gamma}$. In this regime, $T_{21} \approx 0$ and there is no detectable 21-cm signal.

The 21-cm signal remains undetectable until Cosmic Dawn, when the first sources of light switch on, beginning reionization and emitting Lyman-$\alpha$ (Ly$\alpha$) photons. Through the Wouthuysen-Field effect~\cite{Wouthuysen,Field}, consisting in resonant scattering of Ly$\alpha$ photons which can produce a spin-flip, the spin temperature decouples from the CMB and (re)couples to the gas temperature (which is still much lower than the CMB temperature), so $T_s \approx T_k \ll T_{\gamma}$. In this regime, $T_{21}<0$ once more and the 21-cm signal (re)appears in absorption. At star formation continues, at a certain point Ly$\alpha$ coupling saturates, while at the same time the gas is significantly heated by the UV radiation produced by the stars. At a certain point, the gas temperature (to which the spin temperature is still coupled) will surpass the temperature of the CMB photons, \textit{i.e.} $T_s \approx T_k \gg T_{\gamma}$. In this regime, $T_{21}>0$ and the 21-cm signal is seen for the first time in emission. The signal then reaches a maximum before slowly dying out as reionization ends, by which time any residual 21-cm signal arises only from so-called damped Ly$\alpha$ systems.

So far we discussed the evolution of the 21-cm signal qualitatively. To track the evolution of $T_{21}$ quantitatively from Eq.~(\ref{eq:21}), we need to track the evolution of the spin temperature $T_s$ (which we described qualitatively above) as a function of cosmic time. As shown in~\cite{Zaldarriaga:2003du}, $T_s$ is given by:
\begin{eqnarray}
T_s = \frac{T_{\gamma} + (y_c+ y_{{\rm Ly}\alpha})T_k}{1 + y_c + y_{{\rm Ly}\alpha}}\,.
\label{eq:ts}
\end{eqnarray}
In Eq.~(\ref{eq:ts}) $y_c$ and $y_{Ly\alpha}$ are the coupling coefficients for the collisional hyperfine transition (which couples $T_s$ to $T_k$) and for the hyperfine transition mediated by absorption and re-emission of a Ly$\alpha$ photon, and are given by:
\begin{eqnarray}
y_c = \frac{C_{10}}{A_{10}} \frac{T_{*}}{T_k}\,, \quad y_{Ly\alpha} = \frac{P_{10}}{A_{10}} \frac{T_{*}}{T_{Ly\alpha}}\,,
\label{eq:y}
\end{eqnarray}
where $C_{10}$ is the collisional de-excitation rate for the triplet hyperfine level, $T_{*} = 0.068\,{\rm K}$ is the energy of 21-cm photons, $P_{10} \approx  1.3 \times 10^{-12}J_{-21}\,{\rm s}^{ -1}$ is the indirect de-excitation rate of the triplet level due to absorption of Ly$\alpha$ photons followed by decay to the singlet level, and $J_{-21}$ denotes the Ly$\alpha$ background in units of $10^{-21}\,{\rm erg}\,{\rm s}^{-1}\,{\rm cm}^{-2}\,{\rm Hz}^{-1}\,{\rm sr}^{-1}$, which is determined from the global star formation history before the end of reionization. Notice also that in writing Eq.~(\ref{eq:ts}) we have assumed that $T_{{\rm Ly}\alpha} = T_k$, with $T_{{\rm Ly}\alpha}$ the color temperature of the radiation field, an assumption which is well justified if the medium is optically thick to Ly$\alpha$ photons~\cite{Zaldarriaga:2003du}. Finally, the evolution of the gas kinetic temperature $T_k$ is given by:
\begin{eqnarray}
\frac{dT_k(z)}{dz} = \frac{T_k-T_{\gamma}}{H(z)(1 +z)t_c} + \frac{2T_k}{1+z}\,,
\label{eq:gastemperature}
\end{eqnarray}
where $t_c$ is the Compton heating timescale:
\begin{eqnarray}
t_c = \frac{3 m_e c}{8 \sigma_T a_R T^4_{\gamma}} \left [ \frac{1 + f_{He} + x_e}{x_e} \right ] \,,
\label{eq:compton}
\end{eqnarray}
with $m_{e}$ the electron mass, $\sigma_{T}$ the Thomson scattering cross-section, $a_{R}$ the radiation constant, $f_{He}$ the fractional abundance of helium, and $x_{e}$ the ionization fraction. We track the evolution of the ionization fraction $x_e$ as a function of cosmological parameters numerically using the \texttt{RECFAST} code~\cite{Seager:1999bc}. At the redshifts relevant for the EDGES detection we note that $x_e$ is typically slightly larger than $10^{-4}$.

\subsection{The EDGES detection}
\label{subsec:edges}

Significant experimental effort has been devoted to detecting the global 21-cm signal. In particular, the Experiment to Detect the Global EoR Signature (EDGES) was constructed with the aim of detecting the (second) absorption signal arising at Cosmic Dawn from the (re)coupling of the spin and gas temperatures through the Wouthuysen-Field effect. The EDGES detector uses two low-band instruments, each consisting of a dipole antenna coupled to a radio receiver, and operating at $50-100\,{\rm MHz}$. In February 2018 the EDGES collaboration reported the detection of a flattened absorption profile in the sky-averaged radio spectrum, centered at $78\,{\rm MHz}$ and with full-width half-maximum of $19\,{\rm MHz}$~\cite{Bowman:2018yin}.

The signal detected by EDGES is centered at an equivalent redshift of $z \approx 17$ and spans the range $15 \lesssim z \lesssim 20$. This is consistent with expectations concerning the beginning of Cosmic Dawn. However, the best-fit amplitude of the signal is more than a factor of 2 greater than the largest predictions~\cite{Cohen:2016jbh}. In fact, the signal measured by EDGES translates to the following 99\%~confidence level (C.L.) interval for the brightness temperature at redshift $z \approx 17.2$:
\begin{eqnarray}
T_{21}(z \approx 17.2) \simeq -0.5^{+0.2}_{-0.5}\,{\rm K}\,,
\label{eq:21edges}
\end{eqnarray}
whereas standard expectations set $T_{21} \approx -0.2\,{\rm K}$ at the given redshift. This suggests that the ratio $T_{\gamma}/T_s$ in Eq.~(\ref{eq:21}) should be larger than $15$, whereas the standard scenario sets this value to $7$, indicating that either the gas should be much colder than expected (perhaps due to extra sources of non-adiabatic cooling) or the temperature of CMB photons should be much larger than expected (possibly due to extra sources of radiation which were not accounted for). Another possibility which is evident from Eq.~(\ref{eq:21}) is that the Hubble rate at Cosmic Dawn might be lower than expected.

While concerns about the modelling of EDGES data were raised in~\cite{Hills:2018vyr}, the exciting but somewhat anomalous EDGES result has spurred significant attention in the community, with early work suggesting that scattering between dark matter (DM) and baryons could cool the gas to the extent required to explain the EDGES detection~\cite{Barkana:2018lgd,Munoz:2018pzp,Fialkov:2018xre}. Subsequently, a number of other works have been devoted to proposing alternative explanations for the anomalous EDGES detection, or utilizing such a signal to place constraints on fundamental physics assuming the signal itself is genuine (see e.g.~\cite{Feng:2018rje,Berlin:2018sjs,Barkana:2018cct,Fraser:2018acy,DAmico:2018sxd,Hill:2018lfx,Safarzadeh:2018hhg,Hektor:2018qqw,Slatyer:2018aqg,Mitridate:2018iag,Munoz:2018jwq,Witte:2018itc,Li:2018kzs,Jia:2018csj,Schneider:2018xba,Houston:2018vrf,Wang:2018azy,Xiao:2018jyl,Kovetz:2018zan,Jia:2018mkc,Kovetz:2018zes,Lopez-Honorez:2018ipk,Nebrin:2018vqt,Widmark:2019cut,Li:2019loh} for a very limited list of works in these directions). We shall also follow this approach here: restricting our attention to the unified dark sector models we discussed in Sec.~\ref{sec:unified}, we will study how including the EDGES global 21-cm signal improves constraints on the parameters of such models.

How do the two Chaplygin gas models (GCG and MCG) affect the global 21-cm signal at Cosmic Dawn? We make the simplifying but well-motivated assumption that these models do not alter the microphysics of the 21-cm signal: in other words, the spin structure of the Hydrogen atom (and in particular the hyperfine splitting levels) is unaltered when considering a dark sector described by the GCG and MCG models.~\footnote{This assumption might be broken if we consider models where the unified dark fluid couples non-gravitationally to baryons, and hence might alter the spin structure of the Hydrogen atom. In this work, we adopt an effective approach where the physics of the dark fluid is fully specified by their equations of state, Eq.~(\ref{eq:gcg}) and Eq.~(\ref{eq:mcg}). In the absence of an UV-complete description of the dark fluid (which might involve a string completion) it is then perfectly reasonable to assume that the unified dark fluid does not alter 21-cm microphysics.} We also make the additional assumption that the halo mass function, and correspondingly fraction of mass collapsed in haloes which are able to host star-forming galaxies (which in turn is connected to the strength of the 21-cm absorption signal, due to emission of Ly$\alpha$ photons from the first stars, which couple to the hyperfine 21-cm transition through the Wouthuysen-Field coupling), is unaltered with respect to the standard $\Lambda$CDM case. This assumption is perhaps a little less motivated than the former: if Chaplygin gases are the underlying model for the dark sector, it is possible if not plausible that the process of structure formation and halo collapse might be modified compared to the $\Lambda$CDM case. Ultimately, this is an issue which has to be settled by accurate N-body simulations of the process of structure formation in a Chaplygin gas Universe. To the best of our knowledge, this has not been done so far. Therefore, in the following we make the simplyfying assumption that that the halo mass function is unaltered compared to the $\Lambda$CDM case, and defer a more detailed study of this issue to future work.

With the two assumptions we just discussed above, the Chaplygin gas models only affect the 21-cm signal through changes to the background expansion, and in particular to the Hubble rate $H(z)$. Changing $H(z)$ at cosmic dawn alters both $T_s$ and $T_k$, and correspondingly $T_{21}$, as is clear from Eqs.~(\ref{eq:21},\ref{eq:ts},\ref{eq:gastemperature}). As we argued previously, the peculiarity of Chaplygin gas models (aside from their providing an unified description of DM and DE) is the fact that they feature a period of exotic matter domination between the DM- and DE-dominated epochs. In the original Chaplygin gas model~\cite{Kamenshchik:2001cp}, this is a period of stiff matter domination (\textit{i.e.} with equation of state $w=1$), whereas it can be a soft matter domination period (\textit{i.e.} with equation of state $w<1$) in the GCG and MCG models. At any rate, during such an exotic period the energy density of the dark fluid dilutes more quickly than dust. For instance, during a stiff matter domination epoch $\rho \propto a^{-6}$, and more generally $\rho \propto a^{-3(1+w)}$ during an epoch of soft matter domination (whereas $\rho \propto a^{-3}$ for dust): when $w>0$, the energy density of the Chaplygin gas dilutes more quickly than that of dust and hence the Hubble rate is correspondingly lower. This is of course interesting since lowering the Hubble rate at Cosmic Dawn is one possible way of explaining the anomalously large value of the brightness temperature detected by EDGES, as is clear from Eq.~(\ref{eq:21}).

\section{Datasets and analysis methodology}
\label{sec:data}

In the following, we discuss in more detail the cosmological datasets we employ in our analysis, as well as the methodology used.

We use measurements of Cosmic Microwave Background (CMB) temperature and polarization anisotropies, as well as their cross-correlations, from the \textit{Planck} 2015 data release~\cite{Aghanim:2015xee}. We refer to this dataset as ``CMB''. Note that in doing so we use the full measurements of the spectra from the \textit{Planck} mission, and not a compressed version of the latter. We then consider the measurement of the 21-cm brightness temperature at an effective redshift of $17.2$ by the EDGES collaboration, reported in Eq.~(\ref{eq:21edges}) and extracted from the measured global 21-cm absorption signal. We refer to this dataset as ``EDGES'', and we analyze it in combination with the CMB dataset. In principle we could also consider analyzing the EDGES dataset alone, without combining it with CMB measurements. However, such an approach would face strong limitations. On the one side, the EDGES dataset is only sensitive to a certain combination of cosmological parameters, which would be strongly correlated/degenerate with each other. This would lead to an EDGES-only parameter determination being of limited constraining power, and would (somewhat counterintuitively) lead to very slow MCMC runs. On the other hand, including CMB data considerably improves the determination of all cosmological parameters and helps to improve the convergence of our analysis. For this reason, we have chosen not to analyze the EDGES dataset alone.

Let us now discuss the parameters of the models we consider, used to describe the full CMB temperature and polarization anisotropy spectra, as well as the post-recombination expansion history relevant for the EDGES measurement. The parameters not related to the dark sector are the baryon physical energy density $\Omega_bh^2$, the angular size of the sound horizon at decoupling $\theta_s$, the optical depth to reionization $\tau$, and the amplitude and tilt of the primordial scalar power spectrum $\ln(10^{10}A_{\rm s})$ and $n_{\rm s}$. With regards to the parameters characterizing the dark sector, we consider two additional parameters when studying the generalized Chaplygin gas model, and three additional parameter when studying the modified Chaplygin gas model. Within the GCG case, the two extra parameters characterizing the background evolution of the dark fluid are $B_s$ and $\alpha$ [see Eqs.~(\ref{eq:gcg},\ref{eq:rhogcg})], whereas for the MCG case the three extra parameters are $B_s$, $\alpha$ and $B$ [see Eqs.~(\ref{eq:mcg},\ref{eq:rhomcg})]. Notice that we do not consider the Chaplygin gas parameter $A$ as being a free parameter. The reason is that $A$ is fixed by the closure condition, in other words the requirement that the Universe be flat (and hence $\sum_i \Omega_i=1$). In summary, the parameter space of the GCG model is 7-dimensional, whereas that of the MCG model is 8-dimensional.

We impose flat priors on all cosmological parameters, with prior ranges listed in Tab.~\ref{priors}. In particular, the prior ranges on the parameters characterizing the dark sector are chosen in such a way as to avoid perturbative instabilities, following our discussion in Sec.~\ref{subsec:perturbations} and earlier works~\cite{Xu:2012qx,Xu:2012ca}. To analyze the CMB measurements we make use of the public \textit{Planck} likelihood code~\cite{Aghanim:2015xee}. With regards to the EDGES measurements, we model the likelihood as being a Gaussian in the brightness temperature $T_{21}$ with mean $\overline{T}$ and width $\sigma$:
\begin{eqnarray}
{\cal L}^{\rm EDGES} \propto \exp \left [ -\frac{(T_{21}(\boldsymbol{\theta})-\overline{T})^2}{2\sigma^2} \right ]\,.
\label{eq:likelihoodedges}
\end{eqnarray}
where we take $\overline{T} \approx -0.5\,{\rm K}$ and $\sigma \approx 0.35\,{\rm K}$, with $\sigma$ obtained by averaging the upper and lower error bars in Eq.~(\ref{eq:21edges}). In doing so, we have approximated the probability distribution function for $T_{21}$ as measured by EDGES to be symmetrical. Since cosmological constraints from EDGES are almost completely dominated by the depth of the absorption feature rather than its width, we do not expect this approximation to have a noticeable impact on our final results. In Eq.~(\ref{eq:likelihoodedges}), $\boldsymbol{\theta}$ collectively denotes our set of cosmological parameters (including those characterizing the CG models), whereas by $T_{21}(\boldsymbol{\theta})$ we indicate the theoretical prediction for the brightness temperature as a function of cosmological parameters, computed following the methodology presented in Sec.~\ref{subsec:theory}, and in particular using Eqs.~(\ref{eq:21}-\ref{eq:compton}).

From the above discussion and in particular the discussion towards the end Sec.~\ref{subsec:edges}, it follows that as far as the models we are considering are concerned, the EDGES detection effectively provides a measurement of the expansion rate at $z \approx 17.2$. In this sense, we can effectively view the EDGES detection as providing a new high-redshift point on the Hubble diagram (or, in some way, a new cosmic chronometer point), albeit in tension with the low-redshift part of the diagram, at least within the framework of $\Lambda$CDM. We expect this high-redshift point to be particularly useful in constraining Chaplygin gas models due to their non-standard soft/stiff matter behaviour in between their DM- and DE-like regimes: this non-standard regime might alter the expansion rate around Cosmic Dawn, and consequently could be constrained by the EDGES detection (under the assumption that the detection is genuine and not plagued by unknown systematics).

To sample the posterior distribution of the parameter space we make use of Markov Chain Monte Carlo (MCMC) methods, using the publicly available cosmological MCMC sampler \texttt{CosmoMC}~\cite{Lewis:2002ah}. We examine the convergence of the generated chains using the Gelman-Rubin statistic $R-1$~\cite{gelmanrubin}.

\squeezetable
\begin{table}
\begin{center}
\renewcommand{\arraystretch}{1.4}
\begin{tabular}{|c@{\hspace{1 cm}} cc|}
\hline
\textbf{Parameter} & \textbf{GCG prior} & \textbf{MCG prior}\\
\hline\hline
$\Omega_bh^2$ & $[0.005\,,\,0.1]$ & $[0.005\,,\,0.1]$\\
$\theta_s$ & $[0.5\,,\,10]$ & $[0.5\,,\,10]$\\
$\tau$ & $[0.01\,,\,0.8]$ & $[0.01\,,\,0.8]$\\
$\ln(10^{10}A_{s})$ & $[2\,,\,4]$ & $[2\,,\,4]$\\
$n_s$ & $[0.8\,,\, 1.2]$ & $[0.8\,,\, 1.2]$\\
$B_s$ & $[0\,,\,1]$ & $[0\,,\,1]$\\
$\alpha$ & $[0\,,\,1]$ & $[-1\,,\,1]$\\
$B$ & - & $[-1 \,,\,1]$\\
\hline
\end{tabular}
\end{center}
\caption{Flat priors on the cosmological parameters used in this paper. }
\label{priors}
\end{table}

\section{Results}
\label{sec:results}

We now discuss the observational constraints we have obtained for the generalized and modified Chaplygin gas models, using the data and methodology discussed in Sec.~\ref{sec:data}.

\begin{figure*}
\includegraphics[width=0.6\textwidth]{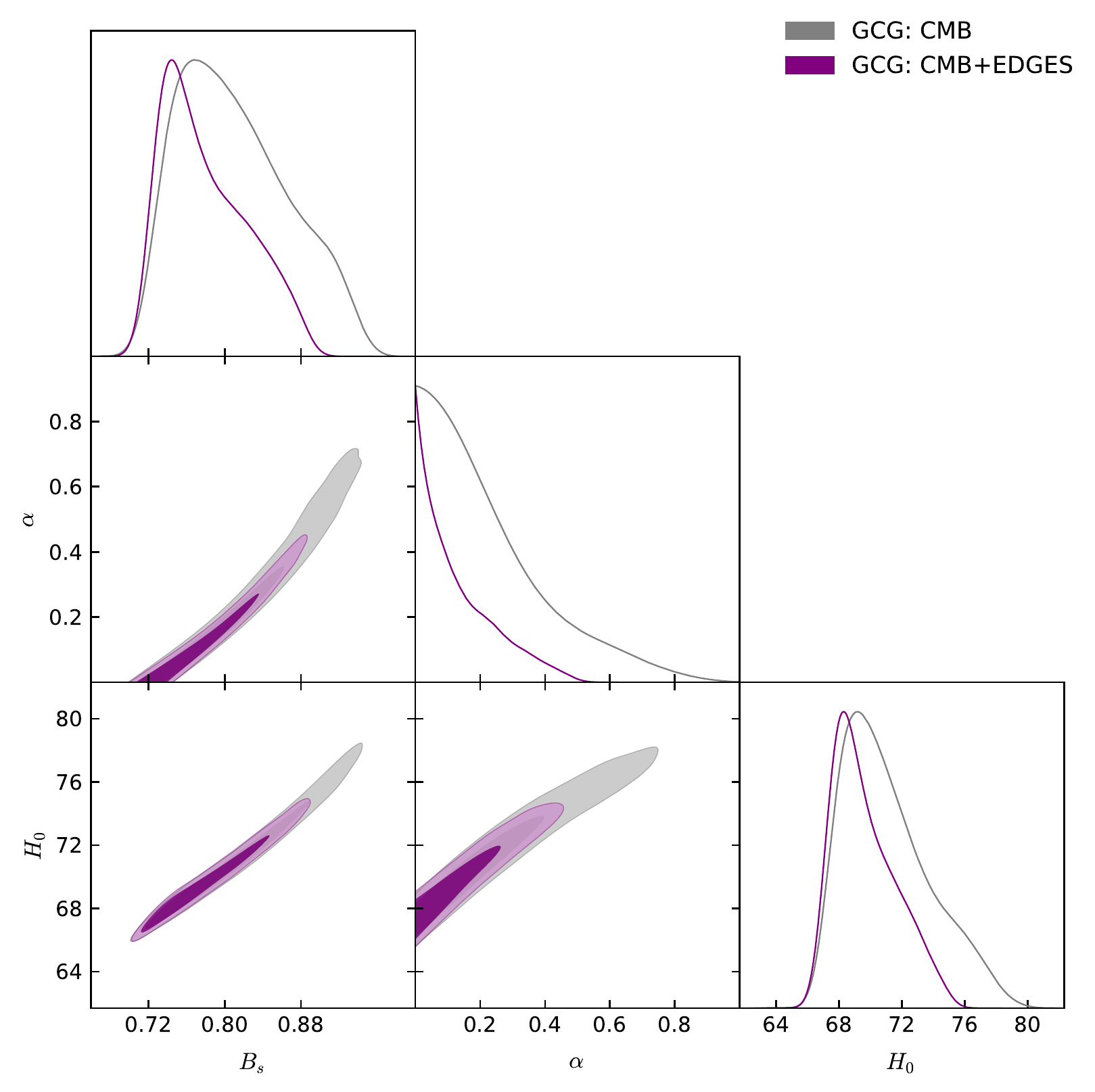}
\caption{Triangle plot showing the 2D joint and 1D marginalized posterior distributions for $B_s$, $\alpha$, and $H_0$ within the GCG model, obtained using only the CMB dataset (grey contours and curves), or the combined CMB+EDGES dataset (purple contours and curves).}
\label{fig:gcg}
\end{figure*}

\begin{figure*}
\includegraphics[width=0.6\textwidth]{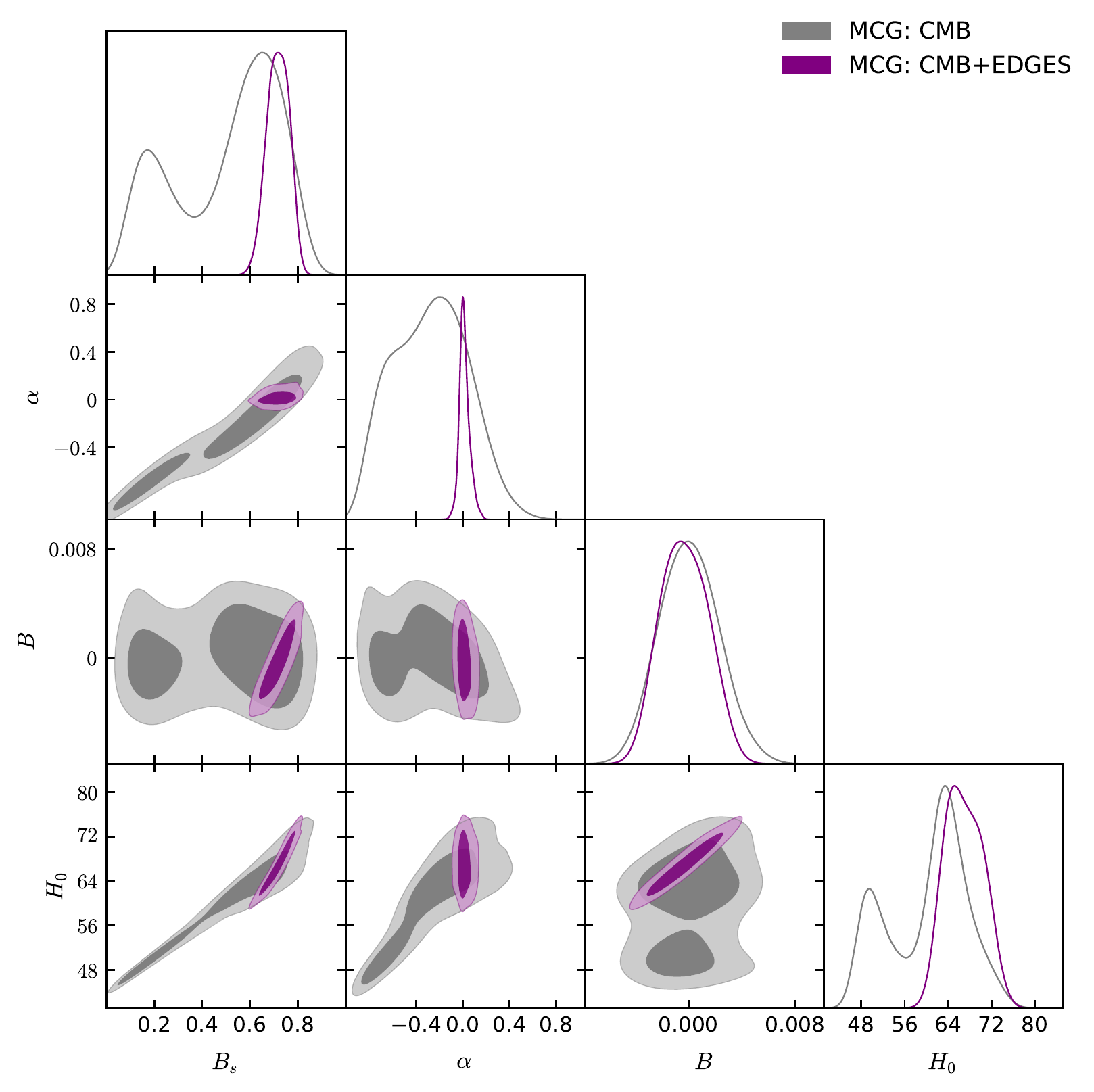}
\caption{As in Fig.~\ref{fig:mcg} but for the modified Chaplygin gas model.}
\label{fig:mcg}
\end{figure*}

In Tab.~\ref{tab:gcg} and Tab.~\ref{tab:mcg} we show the constraints on the main parameters of interest, for both the CMB and CMB+EDGES dataset combinations, displaying either $68\%$~C.L. and $95\%$~C.L. intervals or upper limits depending on the shape of the posterior of the parameter in question. For instance, for the case of $\alpha$ in the GCG model our analysis does not report a detection, but only upper limits, \textit{i.e.} the $68\%$~C.L. and $95\%$~C.L. intervals encompass $\alpha=0$, the lower limit of the prior, which also happens to be where the posterior peaks. In these tables we also report constraints on the Hubble constant $H_0$, although we caution the reader that this is a derived parameter. Finally, in Fig.~\ref{fig:gcg} and Fig.~\ref{fig:mcg} we display triangular plots exhibiting the 2D joint and 1D marginalized posterior distributions for selected parameters of interest ($B_s$, $\alpha$, and $H_0$ for the GCG model; $B_s$, $\alpha$, $B$, and $H_0$ for the MCG model). We now discuss our results in more detail, beginning with the GCG model in Sec.~\ref{subsec:resultsgcg}, before moving on to the MCG model in Sec.~\ref{subsec:resultsmcg}.

\begingroup
\begin{center}
\begin{table*}
\scalebox{1.5}{
\begin{tabular}{ccccccc}
\hline\hline

Parameter & CMB & CMB+EDGES \\ \hline

$\Omega_bh^2$ & $    0.0222_{-    0.0002-    0.0003}^{+    0.0002+    0.0003}$  & $ 0.0222_{-    0.00016-    0.0003}^{+    0.0002+    0.0003}$ \\

$100\theta_s$ & $    1.026_{-    0.001-    0.003}^{+    0.002+    0.002}$ & $    1.027_{-    0.001-    0.002}^{+    0.001+    0.002}$ \\

$\tau$ & $    0.075_{-    0.016-    0.033}^{+    0.017+    0.034}$ & $    0.078_{-    0.017-    0.034}^{+    0.017+    0.033}$ \\

$n_s$ & $    0.965_{-    0.004-    0.009}^{+    0.005+    0.009}$ & $    0.965_{-    0.005-    0.009}^{+    0.005+    0.009}$  \\

$\ln(10^{10} A_s)$ & $    3.09_{-    0.03-    0.06}^{+    0.03+    0.06}$ & 
$    3.09_{-    0.03-    0.07}^{+    0.03+    0.07}$ \\

$B_s$ & $    0.81_{-    0.08-    0.10}^{+    0.04+    0.12}$ & $    0.78_{-    0.06-    0.07}^{+    0.03+    0.09}$  \\

$\alpha$ & $ <0.27\,<0.62$ & $ <0.17\,<0.37 $ \\

$H_0 ({\rm km}\,{\rm s}^{-1}\,{\rm Mpc}^{-1})$ & $   71.0_{-    3.7-    4.6}^{+    1.7+    5.9}$ & $   69.6_{-    2.7-    3.4}^{+    1.4+    4.3}$  \\

\hline\hline
\end{tabular}
}                                                                                           \caption{$68\%$~C.L. and $95\%$~C.L. intervals on the parameters of the generalized Chaplygin gas model obtained by analysing only \textit{Planck} Cosmic Microwave Background data (``CMB'' column) or \textit{Planck} CMB data in combination with the global 21-cm absorption signal measured by EDGES (``CMB+EDGES'' column). For the parameter $\alpha$ we report $68\%$~C.L. and $95\%$~C.L. upper limits, since the posterior distribution of this parameter is not consistent with a detection of non-zero $\alpha$ (\textit{i.e.} the $68\%$~C.L. and $95\%$~C.L. intervals for $\alpha$ encompass $\alpha=0$, the lower limit of the prior). Note that the Hubble constant $H_0$ is a derived parameter.}
\label{tab:gcg}
\end{table*}                       
\end{center}                        
\endgroup

\subsection{Results for the generalized Chaplygin gas model}
\label{subsec:resultsgcg}

In Tab.~\ref{tab:gcg} we show the constraints we obtain on the parameters of the generalized Chaplygin gas model employing both the CMB and CMB+EDGES datasets, whereas in Fig.~\ref{fig:gcg} we display the 2D joint and 1D marginalized posterior distributions for selected parameters ($B_s$, $\alpha$, and $H_0$), for the same dataset combinations (CMB-only in grey, CMB+EDGES in purple).

By inspecting Tab.~\ref{tab:gcg} and Fig.~\ref{fig:gcg}, we can visually see that the determination of the GCG-specific parameters improves substantially when adding the EDGES measurement to the CMB dataset. In fact, we notice that the uncertainties on $B_s$ and $\alpha$ shrink by about a factor of $1.5$ when including the EDGES measurement. In particular, the $95\%$~C.L. upper limit on $\alpha$ improves substantially from $0.62$ to $0.37$, whereas the $95\%$~C.L. uncertainty on $B_s$ improves from $0.12$ to $0.09$. From Fig.~\ref{fig:gcg}, we also see that the parameters $B_s$, $\alpha$, and $H_0$ are quite strongly correlated with each other. The degeneracies between these parameters present with CMB data alone remain essentially unchanged both in strength and direction even after the addition of the EDGES data, which are thus unable to lift these degeneracies. Finally, EDGES data do not improve limits on parameters of parameters such as $\theta_s$, $\tau$, $n_s$, $A_s$, and $\Omega_bh^2$, whose determination is almost exclusively driven by CMB data as one could correctly have expected.

The fact that the EDGES measurement, consisting of only one point with a rather large error bar, has improved the CMB-only estimation of cosmological parameters considerably, might be at first glance surprising. To explain this, we remind the reader that the CMB alone only probes angular fluctuations at last-scattering (which alone are already a remarkable probe of cosmological parameters), resulting from the projection of a physical scale from last-scattering to us. In order to improve the CMB-only determination of cosmological parameters, one needs to be able to extract physical scales out of the observed angular scales. To do so requires distance measurements between last-scattering and today. This is why even a single measurement of the distance-redshift relation (for instance a BAO distance measurement) allows one to considerably improve the CMB-only determination of cosmological parameters. In fact, such measurements essentially help in fixing the angular diameter distance to last-scattering. As we explained earlier in Sec.~\ref{sec:data}, within the MCG and GCG models we can effectively view the EDGES measurement as a new high-redshift point on the Hubble diagram. This measurement of the distance-redshift relation at high redshift helps calibrate the angular fluctuations in the CMB to a physical scale, improving the determination of cosmological parameters from the latter. Moreover, as we discussed earlier, EDGES might be uniquely placed to probe dark sector components which are either non-standard or exhibit non-standard behaviour at high redshift, as in the case of Chaplygin gases due to their intermediate soft matter behaviour. For these two reasons explained above, the EDGES measurement is particularly useful in improving constraints on cosmological parameters within the Chaplygin gas models.

The central value of $B_s \approx 0.8$ is fully consistent with expectations. In fact, from Eq.~(\ref{eq:eosgcg}) we see that the EoS of the GCG fluid today is given by $w_{\rm gcg}=-B_s$. Within the $\Lambda$CDM model, the effective EoS today (given by the weighted average of the EoS of each component, with weights given by the density parameters of the components, with the weighted average hence being dominated by DM and DE) lies between $-0.7$ and $-0.8$, which is consistent with the value of $B_s$ we infer.

The constraints we derive on $H_0$ within the GCG model are rather interesting and deserve a further comment. From the CMB-only dataset, we find $H_0 = 71.0^{+1.7}_{-3.7}\,{\rm km}\,{\rm s}^{-1}\,{\rm Mpc}^{-1}$ at $68\%$~C.L., which is significantly higher than the same value derived within $\Lambda$CDM with the same dataset, for which $H_0 = 67.7 \pm 0.7\,{\rm km}\,{\rm s}^{-1}\,{\rm Mpc}^{-1}$~\cite{Ade:2015xua}. While it is true that the error bar on $H_0$ within the GCG model is larger than the error bar within $\Lambda$CDM, it is also rather noteworthy that the central value of $H_0$ is considerably shifted with respect to the $\Lambda$CDM determination. With this in mind, we see that the tension with the local value of $H_0$, whose most up-to-date estimate from the \textit{Hubble Space Telescope} yields $H_0 = 74.03 \pm 1.42\,{\rm km}\,{\rm s}^{-1}\,{\rm Mpc}^{-1}$~\cite{Riess:2019cxk}, is reduced from $\approx 4\sigma$ down to barely $1.3\sigma$. The reason is that during the exotic soft matter domination period, the energy density of the dark fluid dilutes faster than matter, and hence the expansion rate slows down with respect to that of a DM dominated Universe. In order to then keep the distance to last scattering (and hence $\theta_s$) fixed, it is necessary to increase $H_0$ (see e.g.~\cite{Vagnozzi:2019ezj}). On the other hand, as we saw earlier in Sec.~\ref{subsec:edges}, the EDGES measurement prefer a lower expansion rate, explaining why including EDGES data slightly lowers $H_0$ with respect to the CMB-only determination.

In conclusion, we have found that the EDGES dataset has noticeably improved limits on the parameters characterizing the generalized Chaplygin gas model. In particular, the uncertainties on the parameters $B_s$ and $\alpha$ have improved with respect to the CMB-only case by a factor of $\approx 1.5$. Moreover, we have found that the generalized Chaplygin gas model has the potential to significantly reduce the $H_0$ tension. In light of these results, the generalized Chaplygin gas model still appears to be a viable model for the dark sector. At the same time, it also emerges as an interesting model which has the potential to soften the existing tensions between the concordance $\Lambda$CDM model and the EDGES measurement, as well as with the local distance ladder measurement of the Hubble constant $H_0$. This certainly warrants more detailed studies on the matter, for instance including additional low-redshift measurements, which we plan to address in future work.

\begingroup
\squeezetable
\begin{center}
\begin{table*}
\scalebox{1.5}{
\begin{tabular}{cccccccccccc}
\hline\hline

Parameter & CMB & CMB+EDGES \\ \hline
$\Omega_bh^2$ & $    0.0222_{-    0.0002-    0.0003}^{+    0.0002+    0.0003}$ & $    0.0222_{-    0.0002-    0.0004}^{+    0.0002+    0.0004}$  \\

$100\theta_s$ & $    1.031_{-    0.004-    0.006}^{+    0.006+    0.007}$ & $    1.028_{-    0.002-    0.003}^{+    0.002+    0.003}$ \\

$\tau$ & $    0.077_{-    0.016-    0.034}^{+    0.017+    0.032}$ & $    0.079_{-    0.016-    0.033}^{+    0.017+    0.032}$ \\

$n_s$ & $    0.965_{-    0.007-    0.013}^{+    0.006+    0.013}$ & $    0.965_{-    0.007-    0.011}^{+    0.006+    0.012}$ \\

$\ln(10^{10} A_s)$ & $    3.09_{-    0.03-    0.07}^{+    0.03+    0.06}$ & $    3.09_{-    0.03-    0.07}^{+    0.03+    0.06}$ \\

$B_s$ & $    0.50_{-    0.38-    0.41}^{+    0.30+    0.32}$ & $    0.72_{-    0.05-    0.09}^{+    0.05+    0.09}$ \\

$\alpha$ & $   -0.28_{-    0.38-    0.59}^{+    0.33+    0.57}$ & $    0.02_{-    0.05-    0.09}^{+    0.03+    0.10}$ \\

$B$ & $    0.000_{-    0.002-    0.005}^{+    0.002+    0.004}$ & $   -0.000_{-    0.002-    0.003}^{+    0.002+    0.003}$ \\

$H_0$ & $   60_{-    12-    15}^{+    9+   13}$ & $   66.9_{-    4.1-    6.6}^{+    3.6+    6.9}$ \\

\hline\hline
\end{tabular}
}          
\caption{As in Tab.~\ref{tab:gcg} but for the modified Chaplygin gas model.}
\label{tab:mcg}
\end{table*}                         
\end{center}
\endgroup

\subsection{Results for the modified Chaplygin gas model}
\label{subsec:resultsmcg}

In Tab.~\ref{tab:mcg} we show the constraints we obtain on the parameters of the modified Chaplygin gas model employing both the CMB and CMB+EDGES datasets, whereas in Fig.~\ref{fig:mcg} we display the 2D joint and 1D marginalized posterior distributions for selected parameters ($B_s$, $\alpha$, $B$, and $H_0$), for the same dataset combinations (CMB-only in grey, CMB+EDGES in purple). Recall that, as per our discussion in Sec.~\ref{subsec:perturbations}, within the MCG model $\alpha$ is allowed to take negative values between $-1$ and $0$ (region which instead is not allowed within the GCG model).

By inspecting Tab.~\ref{tab:mcg} and Fig.~\ref{fig:mcg}, we reach conclusions similar to those previously drawn for the GCG model. However, unlike the GCG model, for the case where we only use CMB data, we observe bimodal posterior distributions for $B_s$ and $H_0$, finding additionally that the two are strongly correlated. Again, we find that EDGES data do not improve the limits on parameters such as $\theta_s$, $\tau$, $n_s$, $A_s$, and $\Omega_bh^2$, whose determination is almost exclusively driven by CMB data. On the other hand, the determination of $B_s$ and $\alpha$ is considerably improved by the inclusion of EDGES data, as the latter selects one of the two peaks of the bimodal distribution of $B_s$, and results in the error bars for these two parameters shrinking by a factor between $6$ and $10$. One can also notice that the the inclusion of EDGES data improves the determination of $H_0$ by selecting the higher $H_0$ peak of the posterior distribution.

However, we find that the inclusion of EDGES data does not improve limits on $B$, which remains consistent with $0$ with error bars essentially unaltered. It is also worth noting that, unlike in the GCG case, EDGES data is able to lift degeneracies involving $\alpha$, in particular cutting out the region of negative $\alpha$. In fact, when using only CMB data we find a central value of $\alpha=-0.28$, whereas including EDGES data we find $\alpha=0.02$. As in the GCG case, the values of $B_s$ and $B$ we find are consistent with expectations (the EoS of the MCG fluid today is given by $w_{\rm mcg} = B-B_s$, which is $\approx -0.5$ for the central values of $B \approx 0$ and $B_s \approx 0.5$ we find).

The reason why EDGES data cut the $\alpha<0$ region of parameter space is that negative values of $\alpha$ would actually lead to a higher expansion rate during Cosmic Dawn, inconsistent with the anomalous EDGES detection which instead prefer a lower expansion rate. In fact, this is also the reason why the CMB-only value of $H_0 = 60\,{\rm km}\,{\rm s}^{-1}\,{\rm Mpc}^{-1}$ is, perhaps surprisingly, lower than the $\Lambda$CDM determination, which is the opposite of what happened for the GCG model. A lower value of $H_0$ is required to compensate for the higher expansion rate in the past occurring when $\alpha<0$, and keep the distance to last scattering (and hence $\theta_s$) fixed. The error bar on $H_0$ for the CMB-only dataset is a factor of $12$ larger than the $\Lambda$CDM error bar, implying that the measurement is formally consistent with the local measurement in~\cite{Riess:2019cxk}. However, given the huge error bar involved, we refrain from claiming that the MCG model can solve the $H_0$ tension, especially since the central value of $H_0$ has actually shifted downwards with respect to the $\Lambda$CDM value. On the other hand, the addition of EDGES data cuts the $\alpha<0$ region of parameter space as discussed, and thus slightly raises the value of $H_0$ (while also reducing the error bar).

In conclusion, we have found that also for the modified Chaplygin gas model the inclusion of EDGES data has helped improve limits on $B_s$ and $\alpha$, reducing their uncertainties by a factor as large as $10$, and selecting specific peaks in the posterior distributions of parameters which would otherwise be bimodal when using CMB data alone. In particular, EDGES data help cutting the region of parameter space where $\alpha<0$, allowed by stability considerations but inconsistent with the lower expansion rate at Cosmic Dawn required to explain the EDGES measurement. Thus also the MCG model appears to be a viable cosmological model, despite not being able to satisfactorily solve the $H_0$ tension (although we remark that the local distance ladder estimate of $H_0$ is formally in agreement with the MCG estimate of the latter within better than $2\sigma$).

\section{Conclusions}
\label{sec:conclusions}

Despite a wealth of available extraordinarily precise cosmological datasets, the nature and origin of the dark sector of the Universe remains a mystery to date. While usually treated separately, it is possible that dark matter and dark energy might simply be two manifestations of the same underlying entity, an approach advocated by unified dark sector models. Chaplygin gas models represent an interesting possibility in this sense, as they provide an unified fluid interpolating between an effective dark matter component in the past and an effective dark energy component at present.

The presently available host of exquisitely precise cosmological datasets has recently been augmented by the detection of the global 21-cm absorption signal at Cosmic Dawn by the EDGES experiment. This signal is generated by Ly$\alpha$ radiation emitted from the first stars, coupling the spin temperature to the gas temperature (which at the time is much colder than the temperature of CMB photons) through the Wouthuysen-Field effect. While controversial, the EDGES measurement is particularly suited for tests of non-standard dark sector components (e.g. exotic interactions between dark matter and baryons, or dark matter and dark energy). In this work, we have explored whether the global 21-cm signal detected by EDGES can improve our understanding of Chaplygin gas models of the dark sector.

We have focused on two extensions of the original Chaplygin gas model~\cite{Kamenshchik:2001cp}: the generalized Chaplygin gas~\cite{Bento:2002ps} and the modified Chaplygin gas~\cite{Benaoum:2002zs}. We have first computed constraints on the free parameters of the model using only Cosmic Microwave Background data from the \textit{Planck} satellite, before also including the EDGES measurement. We have found that including the latter considerably improves the determination of parameters characterizing the two Chaplygin gas models, reducing their uncertainty by a factor between $1.5$ and $7$. Our results also suggest that the generalized Chaplygin gas model is an interesting candidate for addressing the tension between CMB and local estimates of the Hubble constant $H_0$ (see Tab.~\ref{tab:gcg}, ``CMB'' column). Both models, in any case, appear to be viable and interesting candidates for the dark sector of the Universe in light of precision cosmological data, and might also shed light on some of the tensions plaguing the $\Lambda$CDM model, namely concerning the EDGES measurement of the global 21-cm signal at Cosmic Dawn, and the local distance ladder measurement of the Hubble constant.

It might be interesting to extend our analysis by forecasting how future CMB missions (such as \textit{Simons Observatory}~\cite{Ade:2018sbj,Abitbol:2019nhf} and CMB-S4~\cite{Abazajian:2016yjj}) or HI intensity mapping surveys such as SKA (following for instance~\cite{Sprenger:2018tdb,Brinckmann:2018owf}) might improve our understanding of Chaplygin gas models as candidates for the dark sector of the Universe, or to extend our analysis to other candidate unified dark sector models or more generally to modified gravity models.

In conclusion, we have demonstrated that the global 21-cm absorption signal detected by EDGES provides not only an extraordinary window onto Cosmic Dawn, but also potentially on models attempting to provide an unified description of the dark sector of the Universe. The era of 21-cm cosmology has just begun and we can only wait to see how future experimental and theoretical efforts in this direction will help shed light on the nature of dark matter and dark energy.

\begin{acknowledgments}
\vskip -0.5 cm
The authors thank the referee for some useful comments that resulted in a significantly improved version. 
W.Y. acknowledges support from the National Natural Science Foundation of China under Grants No. 11705079 and No. 11647153. S.P. acknowledges support from the Mathematical Research Impact Centric Support (MATRICS), File Number: MTR/2018/000940, from the Science and Engineering Research Board, Government of India, as well as from the Faculty Research and Professional Development Fund (FRPDF) Scheme of Presidency University, Kolkata, India. S.V. is supported by the Isaac Newton Trust and the Kavli Foundation through a Newton-Kavli fellowship, and acknowledges a College Research Associateship at Homerton College, University of Cambridge. E.D.V. acknowledges support from the European Research Council in the form of a Consolidator Grant with number 681431. D.F.M. acknowledges support from the Research Council of Norway. S.C. acknowledges support from Istituto Nazionale di Fisica Nucleare (INFN), iniziative specifiche QGSKY and MOONLIGHT2. This work is partially based upon the COST action CA15117 (CANTATA), supported by COST (European Cooperation in Science and Technology). 
\end{acknowledgments}


\begin{thebibliography}{}

\bibitem{Riess:1998cb}
  A.~G.~Riess {\it et al.} [Supernova Search Team],
  {\it Observational evidence from supernovae for an accelerating universe and a cosmological constant,}
  Astron.\ J.\  {\bf 116} (1998) 1009
  [astro-ph/9805201].
  
\bibitem{Perlmutter:1998np}
  S.~Perlmutter {\it et al.} [Supernova Cosmology Project Collaboration],
  {\it Measurements of Omega and Lambda from 42 high redshift supernovae,}
  Astrophys.\ J.\  {\bf 517} (1999) 565
  [astro-ph/9812133].
  
\bibitem{Ester}
  M.~Demianski, E.~Piedipalumbo, C.~Rubano and P.~Scudellaro,
  {\it High redshift cosmography: new results and implication for dark energy,'}
  Mon.\ Not.\ Roy.\ Astron.\ Soc.\  {\bf 426} (2012) 1396
  [arXiv:1206.7046 [astro-ph.CO]].
  
\bibitem{Alam:2016hwk}
  S.~Alam {\it et al.} [BOSS Collaboration],
  {\it The clustering of galaxies in the completed SDSS-III Baryon Oscillation Spectroscopic Survey: cosmological analysis of the DR12 galaxy sample,}
  Mon.\ Not.\ Roy.\ Astron.\ Soc.\  {\bf 470} (2017) no.3,  2617
  [arXiv:1607.03155 [astro-ph.CO]].
  
\bibitem{Troxel:2017xyo}
  M.~A.~Troxel {\it et al.} [DES Collaboration],
  {\it Dark Energy Survey Year 1 results: Cosmological constraints from cosmic shear,}
  Phys.\ Rev.\ D {\bf 98} (2018) no.4,  043528
  [arXiv:1708.01538 [astro-ph.CO]].
  
\bibitem{Aghanim:2018eyx}
  N.~Aghanim {\it et al.} [Planck Collaboration],
  {\it Planck 2018 results. VI. Cosmological parameters,}
  arXiv:1807.06209 [astro-ph.CO].
  
\bibitem{Dodelson:1993je}
  S.~Dodelson and L.~M.~Widrow,
  {\it Sterile-neutrinos as dark matter,}
  Phys.\ Rev.\ Lett.\  {\bf 72} (1994) 17
  [hep-ph/9303287].
  
\bibitem{Jungman:1995df}
  G.~Jungman, M.~Kamionkowski and K.~Griest,
  {\it Supersymmetric dark matter,}
  Phys.\ Rept.\  {\bf 267} (1996) 195
  [hep-ph/9506380].
  
\bibitem{Cirelli:2005uq}
  M.~Cirelli, N.~Fornengo and A.~Strumia,
  {\it Minimal dark matter,}
  Nucl.\ Phys.\ B {\bf 753} (2006) 178
  [hep-ph/0512090].
  
\bibitem{McDonald:2007ex}
  J.~McDonald,
  {\it Gauge singlet scalars as cold dark matter,}
  Phys.\ Rev.\ D {\bf 50} (1994) 3637
  [hep-ph/0702143 [HEP-PH]].
  
\bibitem{ArkaniHamed:2008qn}
  N.~Arkani-Hamed, D.~P.~Finkbeiner, T.~R.~Slatyer and N.~Weiner,
  {\it A Theory of Dark Matter,}
  Phys.\ Rev.\ D {\bf 79} (2009) 015014
  [arXiv:0810.0713 [hep-ph]].
  
\bibitem{Visinelli:2009zm}
  L.~Visinelli and P.~Gondolo,
  {\it Dark Matter Axions Revisited,}
  Phys.\ Rev.\ D {\bf 80} (2009) 035024
  [arXiv:0903.4377 [astro-ph.CO]].
  
\bibitem{Feng:2009mn}
  J.~L.~Feng, M.~Kaplinghat, H.~Tu and H.~B.~Yu,
  {\it Hidden Charged Dark Matter,}
  JCAP {\bf 0907} (2009) 004
  [arXiv:0905.3039 [hep-ph]].
  
\bibitem{Kaplan:2009de}
  D.~E.~Kaplan, G.~Z.~Krnjaic, K.~R.~Rehermann and C.~M.~Wells,
  {\it Atomic Dark Matter,}
  JCAP {\bf 1005} (2010) 021
  [arXiv:0909.0753 [hep-ph]].
  
\bibitem{Fan:2013yva}
  J.~Fan, A.~Katz, L.~Randall and M.~Reece,
  {\it Double-Disk Dark Matter,}
  Phys.\ Dark Univ.\  {\bf 2} (2013) 139
  [arXiv:1303.1521 [astro-ph.CO]].
  
\bibitem{Petraki:2014uza}
  K.~Petraki, L.~Pearce and A.~Kusenko,
  {\it Self-interacting asymmetric dark matter coupled to a light massive dark photon,}
  JCAP {\bf 1407} (2014) 039
  [arXiv:1403.1077 [hep-ph]].
  
\bibitem{Foot:2014uba}
  R.~Foot and S.~Vagnozzi,
  {\it Dissipative hidden sector dark matter,}
  Phys.\ Rev.\ D {\bf 91} (2015) 023512
  [arXiv:1409.7174 [hep-ph]].
  
\bibitem{Foot:2014osa}
  R.~Foot and S.~Vagnozzi,
  {\it Diurnal modulation signal from dissipative hidden sector dark matter,}
  Phys.\ Lett.\ B {\bf 748} (2015) 61
  [arXiv:1412.0762 [hep-ph]].
  
\bibitem{Foot:2016wvj}
  R.~Foot and S.~Vagnozzi,
  {\it Solving the small-scale structure puzzles with dissipative dark matter,}
  JCAP {\bf 1607} (2016) no.07,  013
  [arXiv:1602.02467 [astro-ph.CO]].
  
\bibitem{Visinelli:2017imh}
  L.~Visinelli,
  {\it Light axion-like dark matter must be present during inflation,}
  Phys.\ Rev.\ D {\bf 96} (2017) no.2,  023013
  [arXiv:1703.08798 [astro-ph.CO]].
  
\bibitem{Visinelli:2017qga}
  L.~Visinelli,
  {\it (Non-) thermal production of WIMPs during kination,}
  Symmetry {\bf 10} (2018) no.11,  546
  [arXiv:1710.11006 [astro-ph.CO]].
  
\bibitem{Tenkanen:2019xzn}
  T.~Tenkanen and L.~Visinelli,
  {\it Axion dark matter from Higgs inflation with an intermediate $H_*$,}
  JCAP {\bf 1908} (2019) 033
  arXiv:1906.11837 [astro-ph.CO].
  
\bibitem{Milgrom:1983ca}
  M.~Milgrom,
  {\it A Modification of the Newtonian dynamics as a possible alternative to the hidden mass hypothesis,}
  Astrophys.\ J.\  {\bf 270} (1983) 365.
  
\bibitem{Capozziello:2006ph}
  S.~Capozziello, V.~F.~Cardone and A.~-Troisi,
  {\it Low surface brightness galaxies rotation curves in the low energy limit of $R^n$ gravity: no need for dark matter?,}
  Mon.\ Not.\ Roy.\ Astron.\ Soc.\  {\bf 375} (2007) 1423
  [astro-ph/0603522].
  
\bibitem{Boehmer:2007kx}
  C.~G.~B\"{o}ehmer, T.~Harko and F.~S.~N.~Lobo,
  {\it Dark matter as a geometric effect in f(R) gravity,}
  Astropart.\ Phys.\  {\bf 29} (2008) 386
  [arXiv:0709.0046 [gr-qc]].
  
\bibitem{Chamseddine:2013kea}
  A.~H.~Chamseddine and V.~Mukhanov,
  {\it Mimetic Dark Matter,}
  JHEP {\bf 1311} (2013) 135
  [arXiv:1308.5410 [astro-ph.CO]].
  
\bibitem{Myrzakulov:2015nqa}
  R.~Myrzakulov, L.~Sebastiani, S.~Vagnozzi and S.~Zerbini,
  {\it Mimetic covariant renormalizable gravity,}
  Fund.\ J.\ Mod.\ Phys.\  {\bf 8} (2015) 119
  [arXiv:1505.03115 [gr-qc]].
  
\bibitem{Myrzakulov:2015kda}
  R.~Myrzakulov, L.~Sebastiani, S.~Vagnozzi and S.~Zerbini,
  {\it Static spherically symmetric solutions in mimetic gravity: rotation curves and wormholes,}
  Class.\ Quant.\ Grav.\  {\bf 33} (2016) no.12,  125005
  [arXiv:1510.02284 [gr-qc]].
  
\bibitem{Luongo}
  P.~K.~S.~Dunsby and O.~Luongo,
  {\it On the theory and applications of modern cosmography,}
  Int.\ J.\ Geom.\ Meth.\ Mod.\ Phys.\  {\bf 13} (2016) no.03,  1630002, 
  [arXiv:1511.06532 [gr-qc]].
  
\bibitem{Rinaldi:2016oqp}
  M.~Rinaldi,
  {\it Mimicking dark matter in Horndeski gravity,}
  Phys.\ Dark Univ.\  {\bf 16} (2017) 14
  [arXiv:1608.03839 [gr-qc]].
  
\bibitem{Verlinde:2016toy}
  E.~P.~Verlinde,
  {\it Emergent Gravity and the Dark Universe,}
  SciPost Phys.\  {\bf 2} (2017) no.3,  016
  [arXiv:1611.02269 [hep-th]].
  
\bibitem{Capozziello:2017rvz}
  S.~Capozziello, P.~Jovanovi\'{c}, V.~B.~Jovanovi\'{c} and D.~Borka,
  {\it Addressing the missing matter problem in galaxies through a new fundamental gravitational radius,}
  JCAP {\bf 1706} (2017) no.06,  044
  [arXiv:1702.03430 [gr-qc]].
  
\bibitem{Vagnozzi:2017ilo}
  S.~Vagnozzi,
  {\it Recovering a MOND-like acceleration law in mimetic gravity,}
  Class.\ Quant.\ Grav.\  {\bf 34} (2017) no.18,  185006
  [arXiv:1708.00603 [gr-qc]].
  
  
\bibitem{Sergey}
  S.~Capozziello, K.~F.~Dialektopoulos and S.~V.~Sushkov,
  {\it Classification of the Horndeski cosmologies via Noether Symmetries,}
  Eur.\ Phys.\ J.\ C {\bf 78} (2018) no.6,  447, 
  [arXiv:1803.01429 [gr-qc]].
  
  \bibitem{deHaro:2018sqw}
  J.~de Haro, L.~Arest\'{e} Sal\'{o} and S.~Pan,
  {\it Limiting curvature mimetic gravity and its relation to Loop Quantum Cosmology,}
  Gen.\ Rel.\ Grav.\  {\bf 51} (2019) no.4,  49
  [arXiv:1803.09653 [gr-qc]].
  
\bibitem{Khalifeh:2019zfi}
  A.~R.~Khalifeh, N.~Bellomo, J.~L.~Bernal and R.~Jimenez,
  {\it Can Dark Matter be Geometry? A Case Study with Mimetic Dark Matter,}
  arXiv:1907.03660 [astro-ph.CO].
  
\bibitem{Ratra:1987rm}
  B.~Ratra and P.~J.~E.~Peebles,
  {\it Cosmological Consequences of a Rolling Homogeneous Scalar Field,}
  Phys.\ Rev.\ D {\bf 37} (1988) 3406.
  
\bibitem{Caldwell:1997ii}
  R.~R.~Caldwell, R.~Dave and P.~J.~Steinhardt,
  {\it Cosmological imprint of an energy component with general equation of state,}
  Phys.\ Rev.\ Lett.\  {\bf 80} (1998) 1582
  [astro-ph/9708069].
  
\bibitem{Zlatev:1998tr}
  I.~Zlatev, L.~M.~Wang and P.~J.~Steinhardt,
  {\it Quintessence, cosmic coincidence, and the cosmological constant,}
  Phys.\ Rev.\ Lett.\  {\bf 82} (1999) 896
  [astro-ph/9807002].
  
\bibitem{Freese:2002sq}
  K.~Freese and M.~Lewis,
  {\it Cardassian expansion: A Model in which the universe is flat, matter dominated, and accelerating,}
  Phys.\ Lett.\ B {\bf 540} (2002) 1
  [astro-ph/0201229].
  
\bibitem{Cai:2009zp}
  Y.~F.~Cai, E.~N.~Saridakis, M.~R.~Setare and J.~Q.~Xia,
  {\it Quintom Cosmology: Theoretical implications and observations,}
  Phys.\ Rept.\  {\bf 493} (2010) 1
  [arXiv:0909.2776 [hep-th]].
  
\bibitem{Hlozek:2014lca}
  R.~Hlozek, D.~Grin, D.~J.~E.~Marsh and P.~G.~Ferreira,
  {\it A search for ultralight axions using precision cosmological data,}
  Phys.\ Rev.\ D {\bf 91} (2015) no.10,  103512
  [arXiv:1410.2896 [astro-ph.CO]].
  
\bibitem{Cognola:2016gjy}
  G.~Cognola, R.~Myrzakulov, L.~Sebastiani, S.~Vagnozzi and S.~Zerbini,
  {\it Covariant Ho\v{r}ava-like and mimetic Horndeski gravity: cosmological solutions and perturbations,}
  Class.\ Quant.\ Grav.\  {\bf 33} (2016) no.22,  225014
  [arXiv:1601.00102 [gr-qc]].
  
\bibitem{Vagnozzi:2018jhn}
  S.~Vagnozzi, S.~Dhawan, M.~Gerbino, K.~Freese, A.~Goobar and O.~Mena,
  {\it Constraints on the sum of the neutrino masses in dynamical dark energy models with $w(z) \geq -1$ are tighter than those obtained in $\Lambda$CDM,}
  Phys.\ Rev.\ D {\bf 98} (2018) no.8,  083501
  [arXiv:1801.08553 [astro-ph.CO]].
  
\bibitem{Casalino:2018tcd}
  A.~Casalino, M.~Rinaldi, L.~Sebastiani and S.~Vagnozzi,
  {\it Mimicking dark matter and dark energy in a mimetic model compatible with GW170817,}
  Phys.\ Dark Univ.\  {\bf 22} (2018) 108
  [arXiv:1803.02620 [gr-qc]].
  
\bibitem{Visinelli:2018utg}
  L.~Visinelli and S.~Vagnozzi,
  {\it Cosmological window onto the string axiverse and the supersymmetry breaking scale,}
  Phys.\ Rev.\ D {\bf 99} (2019) no.6,  063517
  [arXiv:1809.06382 [hep-ph]].
  
\bibitem{DiValentino:2019exe}
  E.~Di Valentino, R.~Z.~Ferreira, L.~Visinelli and U.~Danielsson,
  {\it Late time transitions in the quintessence field and the $H_0$ tension,}
  Phys.\ Dark Univ.\  {\bf 26} (2019) 100385
  arXiv:1906.11255 [astro-ph.CO].
  
\bibitem{Visinelli:2019qqu}
  L.~Visinelli, S.~Vagnozzi and U.~Danielsson,
  {\it Revisiting a negative cosmological constant from low-redshift data,}
  Symmetry {\bf 11} (2019) no.8,  1035
  [arXiv:1907.07953 [astro-ph.CO]].
  
\bibitem{Li:2004rb}
  M.~Li,
  {\it A Model of holographic dark energy,}
  Phys.\ Lett.\ B {\bf 603} (2004) 1
  [hep-th/0403127].
  
\bibitem{Nojiri:2006ri}
  S.~Nojiri and S.~D.~Odintsov,
  {\it Introduction to modified gravity and gravitational alternative for dark energy,}
  eConf C {\bf 0602061} (2006) 06
   [Int.\ J.\ Geom.\ Meth.\ Mod.\ Phys.\  {\bf 4} (2007) 115]
  [hep-th/0601213].
  
\bibitem{Capozziello:2006uv}
  S.~Capozziello, V.~F.~Cardone and A.~Troisi,
  {\it Dark energy and dark matter as curvature effects,}
  JCAP {\bf 0608} (2006) 001
  [astro-ph/0602349].
  
\bibitem{Viability}
  S.~Capozziello, S.~Nojiri, S.~D.~Odintsov and A.~Troisi,
  {\it Cosmological viability of f(R)-gravity as an ideal fluid and its compatibility with a matter dominated phase,}
  Phys.\ Lett.\ B {\bf 639} (2006) 135, 
  [astro-ph/0604431].
  
\bibitem{Report}
  S.~Capozziello and M.~De Laurentis,
  {\it Extended Theories of Gravity,}
  Phys.\ Rept.\  {\bf 509} (2011) 167
  [arXiv:1108.6266 [gr-qc]].
  
\bibitem{Felix}
  M.~De Laurentis and A.~J.~Lopez-Revelles,
  {\it Newtonian, Post Newtonian and Parameterized Post Newtonian limits of f(R, G) gravity,}
  Int.\ J.\ Geom.\ Meth.\ Mod.\ Phys.\  {\bf 11} (2014) 1450082,
  [arXiv:1311.0206 [gr-qc]].
  
\bibitem{Hu:2007nk}
  W.~Hu and I.~Sawicki,
  {\it Models of f(R) Cosmic Acceleration that Evade Solar-System Tests,}
  Phys.\ Rev.\ D {\bf 76} (2007) 064004
  [arXiv:0705.1158 [astro-ph]].
  
\bibitem{Appleby:2007vb}
  S.~A.~Appleby and R.~A.~Battye,
  {\it Do consistent $F(R)$ models mimic General Relativity plus $\Lambda$?,}
  Phys.\ Lett.\ B {\bf 654} (2007) 7
  [arXiv:0705.3199 [astro-ph]].
  
\bibitem{Starobinsky:2007hu}
  A.~A.~Starobinsky,
  {\it Disappearing cosmological constant in f(R) gravity,}
  JETP Lett.\  {\bf 86} (2007) 157
  [arXiv:0706.2041 [astro-ph]].

\bibitem{Cognola:2007zu}
  G.~Cognola, E.~Elizalde, S.~Nojiri, S.~D.~Odintsov, L.~Sebastiani and S.~Zerbini,
  {\it A Class of viable modified f(R) gravities describing inflation and the onset of accelerated expansion,}
  Phys.\ Rev.\ D {\bf 77} (2008) 046009
  [arXiv:0712.4017 [hep-th]].
  
\bibitem{Jhingan:2008ym}
  S.~Jhingan, S.~Nojiri, S.~D.~Odintsov, M.~Sami, I.~Thongkool and S.~Zerbini,
  {\it Phantom and non-phantom dark energy: The Cosmological relevance of non-locally corrected gravity,}
  Phys.\ Lett.\ B {\bf 663} (2008) 424
  [arXiv:0803.2613 [hep-th]].
  
\bibitem{Saridakis:2009bv}
  E.~N.~Saridakis,
  {\it Horava-Lifshitz Dark Energy,}
  Eur.\ Phys.\ J.\ C {\bf 67} (2010) 229
  [arXiv:0905.3532 [hep-th]].
  
\bibitem{Appleby:2009uf}
  S.~A.~Appleby, R.~A.~Battye and A.~A.~Starobinsky,
  {\it Curing singularities in cosmological evolution of F(R) gravity,}
  JCAP {\bf 1006} (2010) 005
  [arXiv:0909.1737 [astro-ph.CO]].
  
\bibitem{Dent:2011zz}
  J.~B.~Dent, S.~Dutta and E.~N.~Saridakis,
  {\it f(T) gravity mimicking dynamical dark energy. Background and perturbation analysis,}
  JCAP {\bf 1101} (2011) 009
  [arXiv:1010.2215 [astro-ph.CO]].
  
\bibitem{Myrzakulov:2015qaa}
  R.~Myrzakulov, L.~Sebastiani and S.~Vagnozzi,
  {\it Inflation in $f(R,\phi )$ -theories and mimetic gravity scenario,}
  Eur.\ Phys.\ J.\ C {\bf 75} (2015) 444
  [arXiv:1504.07984 [gr-qc]].
  
\bibitem{Cai:2015emx}
  Y.~F.~Cai, S.~Capozziello, M.~De Laurentis and E.~N.~Saridakis,
  {\it f(T) teleparallel gravity and cosmology,}
  Rept.\ Prog.\ Phys.\  {\bf 79} (2016) no.10,  106901
  [arXiv:1511.07586 [gr-qc]].
  
\bibitem{Sebastiani:2016ras}
  L.~Sebastiani, S.~Vagnozzi and R.~Myrzakulov,
  {\it Mimetic gravity: a review of recent developments and applications to cosmology and astrophysics,}
  Adv.\ High Energy Phys.\  {\bf 2017} (2017) 3156915
  [arXiv:1612.08661 [gr-qc]].
  
\bibitem{Dutta:2017fjw}
  J.~Dutta, W.~Khyllep, E.~N.~Saridakis, N.~Tamanini and S.~Vagnozzi,
  {\it Cosmological dynamics of mimetic gravity,}
  JCAP {\bf 1802} (2018) 041
  [arXiv:1711.07290 [gr-qc]].
  
\bibitem{Casalino:2018wnc}
  A.~Casalino, M.~Rinaldi, L.~Sebastiani and S.~Vagnozzi,
  {\it Alive and well: mimetic gravity and a higher-order extension in light of GW170817,}
  Class.\ Quant.\ Grav.\  {\bf 36} (2019) no.1,  017001
  [arXiv:1811.06830 [gr-qc]].
  
\bibitem{Amendola:1999er}
  L.~Amendola,
  {\it Coupled quintessence,}
  Phys.\ Rev.\ D {\bf 62}, 043511 (2000)
  [astro-ph/9908023].
  
\bibitem{Barrow:2006hia}
  J.~D.~Barrow and T.~Clifton,
  {\it Cosmologies with energy exchange,}
  Phys.\ Rev.\ D {\bf 73} (2006) 103520
  [gr-qc/0604063].
  
\bibitem{He:2008tn}
  J.~H.~He and B.~Wang,
  {\it Effects of the interaction between dark energy and dark matter on cosmological parameters,}
  JCAP {\bf 0806}, 010 (2008)
  [arXiv:0801.4233 [astro-ph]].
  
\bibitem{Valiviita:2008iv} 
  J.~V\"{a}liviita, E.~Majerotto and R.~Maartens,
  {\it Instability in interacting dark energy and dark matter fluids,}
  JCAP {\bf 0807}, 020 (2008)
  [arXiv:0804.0232 [astro-ph]].

\bibitem{Gavela:2009cy}
  M.~B.~Gavela, D.~Hern\'{a}ndez, L.~Lopez Honorez, O.~Mena and S.~Rigolin,
  {\it Dark coupling,}
  JCAP {\bf 0907} (2009) 034
  [arXiv:0901.1611 [astro-ph.CO]].
  
\bibitem{Martinelli:2010rt}
  M.~Martinelli, L.~Lopez Honorez, A.~Melchiorri and O.~Mena,
  {\it Future CMB cosmological constraints in a dark coupled universe,}
  Phys.\ Rev.\ D {\bf 81} (2010) 103534
  [arXiv:1004.2410 [astro-ph.CO]].
  
\bibitem{Pan:2012ki}
  S.~Pan, S.~Bhattacharya and S.~Chakraborty,
  {\it An analytic model for interacting dark energy and its observational constraints,}
  Mon.\ Not.\ Roy.\ Astron.\ Soc.\  {\bf 452} (2015) no.3,  3038
  [arXiv:1210.0396 [gr-qc]].
  
\bibitem{Yang:2014gza}
  W.~Yang and L.~Xu,
  {\it Cosmological constraints on interacting dark energy with redshift-space distortion after Planck data,}
  Phys.\ Rev.\ D {\bf 89} (2014) no.8,  083517
  [arXiv:1401.1286 [astro-ph.CO]].
  
\bibitem{yang:2014vza}
  W.~Yang and L.~Xu,
  {\it Testing coupled dark energy with large scale structure observation,}
  JCAP {\bf 1408} (2014) 034
  [arXiv:1401.5177 [astro-ph.CO]].
  
\bibitem{Yang:2014hea}
  W.~Yang and L.~Xu,
  {\it Coupled dark energy with perturbed Hubble expansion rate,}
  Phys.\ Rev.\ D {\bf 90} (2014) no.8,  083532
  [arXiv:1409.5533 [astro-ph.CO]].
  
\bibitem{Tamanini:2015iia} 
  N.~Tamanini,
  {\it Phenomenological models of dark energy interacting with dark matter,}
  Phys.\ Rev.\ D {\bf 92}, no. 4, 043524 (2015)
  [arXiv:1504.07397 [gr-qc]].
  
\bibitem{Murgia:2016ccp} 
  R.~Murgia, S.~Gariazzo and N.~Fornengo,
  {\it Constraints on the Coupling between Dark Energy and Dark Matter from CMB data,}
  JCAP {\bf 1604}, no. 04, 014 (2016)
  [arXiv:1602.01765 [astro-ph.CO]].

\bibitem{Nunes:2016dlj}
  R.~C.~Nunes, S.~Pan and E.~N.~Saridakis,
  {\it New constraints on interacting dark energy from cosmic chronometers,}
  Phys.\ Rev.\ D {\bf 94} (2016) no.2,  023508
  [arXiv:1605.01712 [astro-ph.CO]].
  
\bibitem{Yang:2016evp}
  W.~Yang, H.~Li, Y.~Wu and J.~Lu,
  {\it Cosmological constraints on coupled dark energy,}
  JCAP {\bf 1610} (2016) no.10,  007
  [arXiv:1608.07039 [astro-ph.CO]].
  
\bibitem{Pan:2016ngu}
  S.~Pan and G.~S.~Sharov,
  {\it A model with interaction of dark components and recent observational data,}
  Mon.\ Not.\ Roy.\ Astron.\ Soc.\  {\bf 472} (2017) no.4,  4736
  [arXiv:1609.02287 [gr-qc]].
  
\bibitem{Shafieloo:2016bpk}
  A.~Shafieloo, D.~K.~Hazra, V.~Sahni and A.~A.~Starobinsky,
  {\it Metastable Dark Energy with Radioactive-like Decay,}
  Mon.\ Not.\ Roy.\ Astron.\ Soc.\  {\bf 473} (2018) no.2,  2760
  [arXiv:1610.05192 [astro-ph.CO]].
  
\bibitem{Sharov:2017iue}
  G.~S.~Sharov, S.~Bhattacharya, S.~Pan, R.~C.~Nunes and S.~Chakraborty,
  {\it A new interacting two fluid model and its consequences,}
  Mon.\ Not.\ Roy.\ Astron.\ Soc.\  {\bf 466} (2017) no.3,  3497
  [arXiv:1701.00780 [gr-qc]].
  
\bibitem{Kumar:2017dnp}
  S.~Kumar and R.~C.~Nunes,
  {\it Echo of interactions in the dark sector,}
  Phys.\ Rev.\ D {\bf 96} (2017) no.10,  103511
  [arXiv:1702.02143 [astro-ph.CO]].
  
\bibitem{DiValentino:2017iww} 
  E.~Di Valentino, A.~Melchiorri and O.~Mena,
  {\it Can interacting dark energy solve the $H_0$ tension?,}
  Phys.\ Rev.\ D {\bf 96}, no. 4, 043503 (2017)
  [arXiv:1704.08342 [astro-ph.CO]].
  
\bibitem{Yang:2017yme}
  W.~Yang, N.~Banerjee and S.~Pan,
  {\it Constraining a dark matter and dark energy interaction scenario with a dynamical equation of state,}
  Phys.\ Rev.\ D {\bf 95} (2017) no.12,  123527
  [arXiv:1705.09278 [astro-ph.CO]].
  
\bibitem{Yang:2017ccc}
  W.~Yang, S.~Pan and D.~F.~Mota,
  {\it Novel approach toward the large-scale stable interacting dark-energy models and their astronomical bounds,}
  Phys.\ Rev.\ D {\bf 96} (2017) no.12,  123508
  [arXiv:1709.00006 [astro-ph.CO]].
  
\bibitem{Yang:2017zjs}
  W.~Yang, S.~Pan and J.~D.~Barrow,
  {\it Large-scale Stability and Astronomical Constraints for Coupled Dark-Energy Models,}
  Phys.\ Rev.\ D {\bf 97} (2018) no.4,  043529
  [arXiv:1706.04953 [astro-ph.CO]].
  
\bibitem{Pan:2017ent}
  S.~Pan, A.~Mukherjee and N.~Banerjee,
  {\it Astronomical bounds on a cosmological model allowing a general interaction in the dark sector,}
  Mon.\ Not.\ Roy.\ Astron.\ Soc.\  {\bf 477} (2018) no.1,  1189
  [arXiv:1710.03725 [astro-ph.CO]].
  
\bibitem{Yang:2018euj}
  W.~Yang, S.~Pan, E.~Di Valentino, R.~C.~Nunes, S.~Vagnozzi and D.~F.~Mota,
  {\it Tale of stable interacting dark energy, observational signatures, and the $H_0$ tension,}
  JCAP {\bf 1809} (2018) no.09,  019
  [arXiv:1805.08252 [astro-ph.CO]].
  
  \bibitem{Yang:2018ubt}
  W.~Yang, S.~Pan, L.~Xu and D.~F.~Mota,
  {\it Effects of anisotropic stress in interacting dark matter – dark energy scenarios,}
  Mon.\ Not.\ Roy.\ Astron.\ Soc.\  {\bf 482} (2019) no.2,  1858
  [arXiv:1804.08455 [astro-ph.CO]].
 
\bibitem{Yang:2018pej}
  W.~Yang, S.~Pan and A.~Paliathanasis,
  {\it Cosmological constraints on an exponential interaction in the dark sector,}
  Mon.\ Not.\ Roy.\ Astron.\ Soc.\  {\bf 482} (2019) no.1,  1007
  [arXiv:1804.08558 [gr-qc]].
  
\bibitem{Yang:2018xlt}
  W.~Yang, S.~Pan, R.~Herrera and S.~Chakraborty,
  {\it Large-scale (in) stability analysis of an exactly solved coupled dark-energy model,}
  Phys.\ Rev.\ D {\bf 98} (2018) no.4,  043517
  [arXiv:1808.01669 [gr-qc]].
  
\bibitem{Yang:2018uae}
  W.~Yang, A.~Mukherjee, E.~Di Valentino and S.~Pan,
  {\it Interacting dark energy with time varying equation of state and the $H_0$ tension,}
  Phys.\ Rev.\ D {\bf 98} (2018) no.12,  123527
  [arXiv:1809.06883 [astro-ph.CO]].
  
\bibitem{Yang:2018qec}
  W.~Yang, N.~Banerjee, A.~Paliathanasis and S.~Pan,
  {\it Reconstructing the dark matter and dark energy interaction scenarios from observations,}
  Phys.\ Dark Univ.\  {\bf 26} (2019) 100383
  arXiv:1812.06854 [astro-ph.CO].
  
\bibitem{Paliathanasis:2019hbi}
  A.~Paliathanasis, S.~Pan and W.~Yang,
  {\it Dynamics of nonlinear interacting dark energy models,}
  Int.\ J.\ Mod.\ Phys.\ D {\bf 28} (2019) no.12,  1950161
  arXiv:1903.02370 [gr-qc].
  
\bibitem{Pan:2019jqh}
  S.~Pan, W.~Yang, C.~Singha and E.~N.~Saridakis,
  {\it Observational constraints on sign-changeable interaction models and alleviation of the $H_0$ tension,}
  Phys.\ Rev.\ D {\bf 100} (2019) no.8,  083539
  arXiv:1903.10969 [astro-ph.CO].
  
\bibitem{Li:2019san}
  X.~L.~Li, A.~Shafieloo, V.~Sahni and A.~A.~Starobinsky,
  {\it Revisiting Metastable Dark Energy and Tensions in the Estimation of Cosmological Parameters,}
  arXiv:1904.03790 [astro-ph.CO].
  
\bibitem{Yang:2019bpr}
  W.~Yang, S.~Pan, E.~Di Valentino, B.~Wang and A.~Wang,
  {\it Forecasting Interacting Vacuum-Energy Models using Gravitational Waves,}
  arXiv:1904.11980 [astro-ph.CO].
  
\bibitem{Yang:2019vni}
  W.~Yang, S.~Vagnozzi, E.~Di Valentino, R.~C.~Nunes, S.~Pan and D.~F.~Mota,
  {\it Listening to the sound of dark sector interactions with gravitational wave standard sirens,}
  JCAP {\bf 1907} (2019) 037
  arXiv:1905.08286 [astro-ph.CO].
  
\bibitem{Yang:2019uzo}
  W.~Yang, O.~Mena, S.~Pan and E.~Di Valentino,
  {\it Dark sectors with dynamical coupling,}
  Phys.\ Rev.\ D {\bf 100} (2019) no.8,  083509
  arXiv:1906.11697 [astro-ph.CO].
  
\bibitem{DiValentino:2019ffd}
  E.~Di Valentino, A.~Melchiorri, O.~Mena and S.~Vagnozzi,
  {\it Interacting dark energy after the latest Planck, DES, and $H_0$ measurements: an excellent solution to the $H_0$ and cosmic shear tensions,}
  arXiv:1908.04281 [astro-ph.CO].
  
\bibitem{Benetti:2019lxu}
  M.~Benetti, W.~Miranda, H.~A.~Borges, C.~Pigozzo, S.~Carneiro and J.~S.~Alcaniz,
  {\it Looking for interactions in the cosmological dark sector,}
  arXiv:1908.07213 [astro-ph.CO].
  
\bibitem{Mukhopadhyay:2019jla}
  U.~Mukhopadhyay, A.~Paul and D.~Majumdar,
  {\it Probing Pseudo Nambu Goldstone Boson Dark Energy Models with Dark Matter -- Dark Energy Interaction,}
  arXiv:1909.03925 [astro-ph.CO].
  
\bibitem{Carneiro:2019rly}
  S.~Carneiro, H.~A.~Borges, R.~von Marttens, J.~S.~Alcaniz and W.~Zimdahl,
  {\it Unphysical properties in a class of interacting dark energy models,}
  arXiv:1909.10336 [gr-qc].
  
\bibitem{Kase:2019veo}
  R.~Kase and S.~Tsujikawa,
  {\it Scalar-field dark energy nonminimally and kinetically coupled to dark matter,}
  arXiv:1910.02699 [gr-qc].
  
\bibitem{Yang:2019uog}
  W.~Yang, S.~Pan, R.~C.~Nunes and D.~F.~Mota,
  {\it Dark calling Dark: Interaction in the dark sector in presence of neutrino properties after Planck CMB final release,}
  arXiv:1910.08821 [astro-ph.CO].
  
\bibitem{DiValentino:2019jae}
  E.~Di Valentino, A.~Melchiorri, O.~Mena and S.~Vagnozzi,
  {\it Non-minimal dark sector physics and cosmological tensions,}
  arXiv:1910.09853 [astro-ph.CO].

\bibitem{Rocco}
  S.~Capozziello, R.~D'Agostino and O.~Luongo,
  {\it Extended Gravity Cosmography}
  Int.\ J.\ Mod.\ Phys.\ D {\bf 28} (2019) no.10,  1930016
  [arXiv:1904.01427 [gr-qc]].
   
\bibitem{Anton1} 
  S.~Capozziello, R.~D'Agostino, R.~Giamb\'{o} and O.~Luongo,
  {\it Effective field description of the Anton-Schmidt cosmic fluid,}
  Phys.\ Rev.\ D {\bf 99} (2019) 023532, 
  [arXiv:1810.05844 [gr-qc]].
  
 \bibitem{Anton2} 
  S.~Capozziello, R.~D'Agostino and O.~Luongo,
  {\it Cosmic acceleration from a single fluid description,}
  Phys.\ Dark Univ.\  {\bf 20} (2018) 1, 
  [arXiv:1712.04317 [gr-qc]].
  
\bibitem{Kamenshchik:2001cp}
  A.~Y.~Kamenshchik, U.~Moschella and V.~Pasquier,
  {\it An Alternative to quintessence,}
  Phys.\ Lett.\ B {\bf 511} (2001) 265
  [gr-qc/0103004].
  
\bibitem{Chaplygin}
  S. Chaplygin, Sci. Mem. Moscow Univ. Math. Phys. \textbf{21}, 1 (1904).
  
\bibitem{Bordemann:1993ep}
  M.~Bordemann and J.~Hoppe,
  {\it The Dynamics of relativistic membranes. 1. Reduction to two-dimensional fluid dynamics,}
  Phys.\ Lett.\ B {\bf 317} (1993) 315
  [hep-th/9307036].
  
\bibitem{Jackiw:2000cc}
  R.~Jackiw and A.~P.~Polychronakos,
  {\it Supersymmetric fluid mechanics,}
  Phys.\ Rev.\ D {\bf 62} (2000) 085019
  [hep-th/0004083].
  
\bibitem{Bento:2002ps}
  M.~C.~Bento, O.~Bertolami and A.~A.~Sen,
  {\it Generalized Chaplygin gas, accelerated expansion and dark energy matter unification,}
  Phys.\ Rev.\ D {\bf 66} (2002) 043507
  [gr-qc/0202064].
  
\bibitem{Benaoum:2002zs}
  H.~B.~Benaoum,
  {\it Accelerated universe from modified Chaplygin gas and tachyonic fluid,}
  hep-th/0205140.
  
\bibitem{Bilic:2001cg}
  N.~Bilic, G.~B.~Tupper and R.~D.~Viollier,
  {\it Unification of dark matter and dark energy: The Inhomogeneous Chaplygin gas,}
  Phys.\ Lett.\ B {\bf 535} (2002) 17
  [astro-ph/0111325].
  
\bibitem{Dev:2002qa}
  A.~Dev, D.~Jain and J.~S.~Alcaniz,
  {\it Cosmological consequences of a Chaplygin gas dark energy,}
  Phys.\ Rev.\ D {\bf 67} (2003) 023515
  [astro-ph/0209379].
  
\bibitem{Gorini:2002kf}
  V.~Gorini, A.~Kamenshchik and U.~Moschella,
  {\it Can the Chaplygin gas be a plausible model for dark energy?,}
  Phys.\ Rev.\ D {\bf 67} (2003) 063509
  [astro-ph/0209395].
  
\bibitem{Makler:2002jv}
  M.~Makler, S.~Quinet de Oliveira and I.~Waga,
  {\it Constraints on the generalized Chaplygin gas from supernovae observations,}
  Phys.\ Lett.\ B {\bf 555} (2003) 1
  [astro-ph/0209486].
  
\bibitem{Bento:2002yx}
  M.~d.~C.~Bento, O.~Bertolami and A.~A.~Sen,
  {\it Generalized Chaplygin gas and CMBR constraints,}
  Phys.\ Rev.\ D {\bf 67} (2003) 063003
  [astro-ph/0210468].
  
\bibitem{Alcaniz:2002yt}
  J.~S.~Alcaniz, D.~Jain and A.~Dev,
  {\it High - redshift objects and the generalized Chaplygin gas,}
  Phys.\ Rev.\ D {\bf 67} (2003) 043514
  [astro-ph/0210476].

\bibitem{Bento:2003we}
  M.~d.~C.~Bento, O.~Bertolami and A.~A.~Sen,
  {\it WMAP constraints on the generalized Chaplygin gas model,}
  Phys.\ Lett.\ B {\bf 575} (2003) 172
  [astro-ph/0303538].
  
\bibitem{Amendola:2003bz}
  L.~Amendola, F.~Finelli, C.~Burigana and D.~Carturan,
  {\it WMAP and the generalized Chaplygin gas,}
  JCAP {\bf 0307} (2003) 005
  [astro-ph/0304325].
  
\bibitem{Dev:2003cx}
  A.~Dev, D.~Jain and J.~S.~Alcaniz,
  {\it Constraints on Chaplygin quartessence from the CLASS gravitational lens statistics and supernova data,}
  Astron.\ Astrophys.\  {\bf 417} (2004) 847
  [astro-ph/0311056].
  
\bibitem{Bertolami:2004ic}
  O.~Bertolami, A.~A.~Sen, S.~Sen and P.~T.~Silva,
  {\it Latest supernova data in the framework of Generalized Chaplygin Gas model,}
  Mon.\ Not.\ Roy.\ Astron.\ Soc.\  {\bf 353} (2004) 329
  [astro-ph/0402387].
  
\bibitem{Debnath:2004cd}
  U.~Debnath, A.~Banerjee and S.~Chakraborty,
  {\it Role of modified Chaplygin gas in accelerated universe,}
  Class.\ Quant.\ Grav.\  {\bf 21} (2004) 5609
  [gr-qc/0411015].
  
\bibitem{Zhang:2004gc}
  X.~Zhang, F.~Q.~Wu and J.~Zhang,
  {\it A New generalized Chaplygin gas as a scheme for unification of dark energy and dark matter,}
  JCAP {\bf 0601} (2006) 003
  [astro-ph/0411221].
  
\bibitem{Sen:2005sk}
  A.~A.~Sen and R.~J.~Scherrer,
  {\it Generalizing the generalized Chaplygin gas,}
  Phys.\ Rev.\ D {\bf 72} (2005) 063511
  [astro-ph/0507717].
  
\bibitem{Zhang:2005jj}
  H.~S.~Zhang and Z.~H.~Zhu,
  {\it Interacting chaplygin gas,}
  Phys.\ Rev.\ D {\bf 73} (2006) 043518
  [astro-ph/0509895].
  
\bibitem{Wu:2006pe}
  P.~Wu and H.~W.~Yu,
  {\it Generalized Chaplygin gas model: Constraints from Hubble parameter versus redshift data,}
  Phys.\ Lett.\ B {\bf 644} (2007) 16
  [gr-qc/0612055].
  
\bibitem{BouhmadiLopez:2007ts}
  M.~Bouhmadi-Lopez and R.~Lazkoz,
  {\it Chaplygin DGP cosmologies,}
  Phys.\ Lett.\ B {\bf 654} (2007) 51
  [arXiv:0706.3896 [astro-ph]].
  
\bibitem{Gorini:2007ta}
  V.~Gorini, A.~Y.~Kamenshchik, U.~Moschella, O.~F.~Piattella and A.~A.~Starobinsky,
  {\it Gauge-invariant analysis of perturbations in Chaplygin gas unified models of dark matter and dark energy,}
  JCAP {\bf 0802} (2008) 016
  [arXiv:0711.4242 [astro-ph]].
  
\bibitem{Ali:2010sv}
  A.~Ali, S.~Dutta, E.~N.~Saridakis and A.~A.~Sen,
  {\it Horava-Lifshitz cosmology with generalized Chaplygin gas,}
  Gen.\ Rel.\ Grav.\  {\bf 44} (2012) 657
  [arXiv:1004.2474 [astro-ph.CO]].
  
\bibitem{Xu:2010zzb}
  L.~Xu and J.~Lu,
  {\it Cosmological constraints on generalized Chaplygin gas model: Markov Chain Monte Carlo approach,}
  JCAP {\bf 1003} (2010) 025
  [arXiv:1004.3344 [astro-ph.CO]].
  
\bibitem{Lu:2010zzf}
  J.~Lu, Y.~Gui and L.~X.~Xu,
  {\it Observational constraint on generalized Chaplygin gas model,}
  Eur.\ Phys.\ J.\ C {\bf 63} (2009) 349
  [arXiv:1004.3365 [astro-ph.CO]].
  
\bibitem{Xu:2012qx}
  L.~Xu, J.~Lu and Y.~Wang,
  {\it Revisiting Generalized Chaplygin Gas as a Unified Dark Matter and Dark Energy Model,}
  Eur.\ Phys.\ J.\ C {\bf 72} (2012) 1883
  [arXiv:1204.4798 [astro-ph.CO]].
  
\bibitem{Xu:2012ca}
  L.~Xu, Y.~Wang and H.~Noh,
  {\it Modified Chaplygin Gas as a Unified Dark Matter and Dark Energy Model and Cosmic Constraints,}
  Eur.\ Phys.\ J.\ C {\bf 72} (2012) 1931
  [arXiv:1204.5571 [astro-ph.CO]].
  
\bibitem{Campos:2012ez}
  J.~P.~Campos, J.~C.~Fabris, R.~Perez, O.~F.~Piattella and H.~Velten,
  {\it Does Chaplygin gas have salvation?,}
  Eur.\ Phys.\ J.\ C {\bf 73} (2013) no.4,  2357
  [arXiv:1212.4136 [astro-ph.CO]].
  
\bibitem{Wang:2013qy}
  Y.~Wang, D.~Wands, L.~Xu, J.~De-Santiago and A.~Hojjati,
  {\it Cosmological constraints on a decomposed Chaplygin gas,}
  Phys.\ Rev.\ D {\bf 87} (2013) no.8,  083503
  [arXiv:1301.5315 [astro-ph.CO]].
  
\bibitem{Khurshudyan:2014ewa}
  E.~O.~Kahya, M.~Khurshudyan, B.~Pourhassan, R.~Myrzakulov and A.~Pasqua,
  {\it Higher order corrections of the extended Chaplygin gas cosmology with varying $G$ and $\Lambda $,}
  Eur.\ Phys.\ J.\ C {\bf 75} (2015) no.2,  43
  [arXiv:1402.2592 [gr-qc]].
  
\bibitem{Avelino:2014nva}
  P.~P.~Avelino, K.~Bolejko and G.~F.~Lewis,
  {\it Nonlinear Chaplygin Gas Cosmologies,}
  Phys.\ Rev.\ D {\bf 89} (2014) no.10,  103004
  [arXiv:1403.1718 [astro-ph.CO]].
  
\bibitem{Sharov:2015ifa}
  G.~S.~Sharov,
  {\it Observational constraints on cosmological models with Chaplygin gas and quadratic equation of state,}
  JCAP {\bf 1606} (2016) no.06,  023
  [arXiv:1506.05246 [gr-qc]].
  
\bibitem{Khurshudyan:2015mva}
  M.~Khurshudyan and R.~Myrzakulov,
  {\it Phase space analysis of some interacting Chaplygin gas models,}
  Eur.\ Phys.\ J.\ C {\bf 77} (2017) no.2,  65
  [arXiv:1509.02263 [gr-qc]].

\bibitem{vonMarttens:2017njo}
  R.~F.~vom Marttens, L.~Casarini, W.~Zimdahl, W.~S.~Hip\'{o}lito-Ricaldi and D.~F.~Mota,
  {\it Does a generalized Chaplygin gas correctly describe the cosmological dark sector?,}
  Phys.\ Dark Univ.\  {\bf 15} (2017) 114
  [arXiv:1702.00651 [astro-ph.CO]].
  
\bibitem{Yang:2019jwn}
  W.~Yang, S.~Pan, A.~Paliathanasis, S.~Ghosh and Y.~Wu,
  {\it Observational constraints of a new unified dark fluid and the $H_0$ tension,}
  Mon.\ Not.\ Roy.\ Astron.\ Soc.\  {\bf 490} (2019) no.2,  2071
  arXiv:1904.10436 [gr-qc].
  
\bibitem{Sandvik:2002jz}
  H.~Sandvik, M.~Tegmark, M.~Zaldarriaga and I.~Waga,
  {\it The end of unified dark matter?,}
  Phys.\ Rev.\ D {\bf 69} (2004) 123524
  [astro-ph/0212114].
  
\bibitem{Reis:2003mw}
  R.~R.~R.~Reis, I.~Waga, M.~O.~Calvao and S.~E.~Joras,
  {\it Entropy perturbations in quartessence Chaplygin models,}
  Phys.\ Rev.\ D {\bf 68} (2003) 061302
  [astro-ph/0306004].
  
\bibitem{HipolitoRicaldi:2009je}
  W.~S.~Hipolito-Ricaldi, H.~E.~S.~Velten and W.~Zimdahl,
  {\it Non-adiabatic dark fluid cosmology,}
  JCAP {\bf 0906} (2009) 016
  [arXiv:0902.4710 [astro-ph.CO]].
  
\bibitem{Bento:2004uh}
  M.~C.~Bento, O.~Bertolami and A.~A.~Sen,
  {\it The Revival of the unified dark energy - dark matter model?,}
  Phys.\ Rev.\ D {\bf 70} (2004) 083519
  [astro-ph/0407239].
  
\bibitem{HipolitoRicaldi:2010mf}
  W.~S.~Hipolito-Ricaldi, H.~E.~S.~Velten and W.~Zimdahl,
 {\it The Viscous Dark Fluid Universe,}
  Phys.\ Rev.\ D {\bf 82} (2010) 063507
  [arXiv:1007.0675 [astro-ph.CO]].
  
\bibitem{Zimdahl:2005ir}
  W.~Zimdahl and J.~C.~Fabris,
  {\it Chaplygin gas with non-adiabatic pressure perturbations,}
  Class.\ Quant.\ Grav.\  {\bf 22} (2005) 4311
  [gr-qc/0504088].
  
\bibitem{Borges:2013bya}
  H.~A.~Borges, S.~Carneiro, J.~C.~Fabris and W.~Zimdahl,
  {\it Non-adiabatic Chaplygin gas,}
  Phys.\ Lett.\ B {\bf 727} (2013) 37
  [arXiv:1306.0917 [astro-ph.CO]].
  
\bibitem{Carneiro:2014jza}
  S.~Carneiro and C.~Pigozzo,
  {\it Observational tests of non-adiabatic Chaplygin gas,}
  JCAP {\bf 1410} (2014) 060
  [arXiv:1407.7812 [astro-ph.CO]].
  
\bibitem{Wouthuysen}
  S.~A.~Wouthuysen,
  {\it On the excitation mechanism of the 21-cm (radio-frequency) interstellar hydrogen emission line,}
  Astron.\ J.\ {\bf 57} (1952) 31
  
\bibitem{Field}
  G.~B.~Field,
  {\it The Spin Temperature of Intergalactic Neutral Hydrogen,}
  Astrophys.\ J.\ {\bf 129} (1959) 356
  
\bibitem{Hirata:2005mz}
  C.~M.~Hirata,
  {\it Wouthuysen-Field coupling strength and application to high-redshift 21 cm radiation,}
  Mon.\ Not.\ Roy.\ Astron.\ Soc.\  {\bf 367} (2006) 259
  [astro-ph/0507102].
  
\bibitem{Bowman:2018yin} 
  J.~D.~Bowman, A.~E.~E.~Rogers, R.~A.~Monsalve, T.~J.~Mozdzen and N.~Mahesh,
  {\it An absorption profile centred at 78 megahertz in the sky-averaged spectrum,}
  Nature {\bf 555}, no. 7694, 67 (2018)
  [arXiv:1810.05912 [astro-ph.CO]].
  
\bibitem{Palanque-Delabrouille:2015pga}
  N.~Palanque-Delabrouille {\it et al.},
 {\it Neutrino masses and cosmology with Lyman-alpha forest power spectrum,}
  JCAP {\bf 1511} (2015) no.11,  011
  [arXiv:1506.05976 [astro-ph.CO]].
  
\bibitem{Giusarma:2016phn}
  E.~Giusarma, M.~Gerbino, O.~Mena, S.~Vagnozzi, S.~Ho and K.~Freese,
  {\it Improvement of cosmological neutrino mass bounds,}
  Phys.\ Rev.\ D {\bf 94} (2016) no.8,  083522
  [arXiv:1605.04320 [astro-ph.CO]].
  
\bibitem{Vagnozzi:2017ovm}
  S.~Vagnozzi, E.~Giusarma, O.~Mena, K.~Freese, M.~Gerbino, S.~Ho and M.~Lattanzi,
  {\it Unveiling $\nu$ secrets with cosmological data: neutrino masses and mass hierarchy,}
  Phys.\ Rev.\ D {\bf 96} (2017) no.12,  123503
  [arXiv:1701.08172 [astro-ph.CO]].
  
\bibitem{Giusarma:2018jei}
  E.~Giusarma, S.~Vagnozzi, S.~Ho, S.~Ferraro, K.~Freese, R.~Kamen-Rubio and K.~B.~Luk,
  {\it Scale-dependent galaxy bias, CMB lensing-galaxy cross-correlation, and neutrino masses,}
  Phys.\ Rev.\ D {\bf 98} (2018) no.12,  123526
  [arXiv:1802.08694 [astro-ph.CO]].
  
\bibitem{Vagnozzi:2019utt}
  S.~Vagnozzi,
  {\it Cosmological searches for the neutrino mass scale and mass ordering,}
  arXiv:1907.08010 [astro-ph.CO].
  
\bibitem{Simpson:2017qvj}
  F.~Simpson, R.~Jimenez, C.~Pena-Garay and L.~Verde,
  {\it Strong Bayesian Evidence for the Normal Neutrino Hierarchy,}
  JCAP {\bf 1706} (2017) no.06,  029
  [arXiv:1703.03425 [astro-ph.CO]].
  
\bibitem{Schwetz:2017fey}
  T.~Schwetz, K.~Freese, M.~Gerbino, E.~Giusarma, S.~Hannestad, M.~Lattanzi, O.~Mena and S.~Vagnozzi,
  {\it Comment on "Strong Evidence for the Normal Neutrino Hierarchy",}
  arXiv:1703.04585 [astro-ph.CO].

\bibitem{Ma:1995ey}
  C.~P.~Ma and E.~Bertschinger,
  {\it Cosmological perturbation theory in the synchronous and conformal Newtonian gauges,}
  Astrophys.\ J.\  {\bf 455} (1995) 7
  [astro-ph/9506072].

\bibitem{Xu:2012zm}
  L.~Xu,
  {\it Unified Dark Fluid with Constant Adiabatic Sound Speed: Including Entropic Perturbations,}
  Phys.\ Rev.\ D {\bf 87}, no. 4, 043503 (2013)
  [arXiv:1210.7413 [astro-ph.CO]].
  
\bibitem{Furlanetto:2006jb}
  S.~Furlanetto, S.~P.~Oh and F.~Briggs,
  {\it Cosmology at Low Frequencies: The 21 cm Transition and the High-Redshift Universe,}
  Phys.\ Rept.\  {\bf 433} (2006) 181
  [astro-ph/0608032].
  
\bibitem{Pritchard:2011xb}
  J.~R.~Pritchard and A.~Loeb,
  {\it 21-cm cosmology,}
  Rept.\ Prog.\ Phys.\  {\bf 75} (2012) 086901
  [arXiv:1109.6012 [astro-ph.CO]].
  
\bibitem{Barkana:2016nyr}
  R.~Barkana,
  {\it The Rise of the First Stars: Supersonic Streaming, Radiative Feedback, and 21-cm Cosmology,}
  Phys.\ Rept.\  {\bf 645} (2016) 1
  [arXiv:1605.04357 [astro-ph.CO]].

\bibitem{Zaldarriaga:2003du}
  M.~Zaldarriaga, S.~R.~Furlanetto and L.~Hernquist,
  {\it 21 Centimeter fluctuations from cosmic gas at high redshifts,}
  Astrophys.\ J.\  {\bf 608} (2004) 622
  [astro-ph/0311514].

\bibitem{Seager:1999bc}
  S.~Seager, D.~D.~Sasselov and D.~Scott,
  {\it A new calculation of the recombination epoch,}
  Astrophys.\ J.\  {\bf 523} (1999) L1
  [astro-ph/9909275].
  
\bibitem{Cohen:2016jbh}
  A.~Cohen, A.~Fialkov, R.~Barkana and M.~Lotem,
  {\it Charting the Parameter Space of the Global 21-cm Signal,}
  Mon.\ Not.\ Roy.\ Astron.\ Soc.\  {\bf 472} (2017) no.2,  1915
  [arXiv:1609.02312 [astro-ph.CO]].

\bibitem{Hills:2018vyr}
  R.~Hills, G.~Kulkarni, P.~D.~Meerburg and E.~Puchwein,
  {\it Concerns about modelling of the EDGES data,}
  Nature {\bf 564} (2018) no.7736,  E32
  [arXiv:1805.01421 [astro-ph.CO]].
  
\bibitem{Barkana:2018lgd}
  R.~Barkana,
  {\it Possible interaction between baryons and dark-matter particles revealed by the first stars,}
  Nature {\bf 555} (2018) no.7694,  71
  [arXiv:1803.06698 [astro-ph.CO]].
  
\bibitem{Munoz:2018pzp}
  J.~B.~Mu\~{n}oz and A.~Loeb,
  {\it A small amount of mini-charged dark matter could cool the baryons in the early Universe,}
  Nature {\bf 557} (2018) no.7707,  684
  [arXiv:1802.10094 [astro-ph.CO]].
  
\bibitem{Fialkov:2018xre}
  A.~Fialkov, R.~Barkana and A.~Cohen,
  {\it Constraining Baryon--Dark Matter Scattering with the Cosmic Dawn 21-cm Signal,}
  Phys.\ Rev.\ Lett.\  {\bf 121} (2018) 011101
  [arXiv:1802.10577 [astro-ph.CO]].
  
\bibitem{Feng:2018rje}
  C.~Feng and G.~Holder,
  {\it Enhanced global signal of neutral hydrogen due to excess radiation at cosmic dawn,}
  Astrophys.\ J.\  {\bf 858} (2018) no.2,  L17
  [arXiv:1802.07432 [astro-ph.CO]].
  
\bibitem{Berlin:2018sjs}
  A.~Berlin, D.~Hooper, G.~Krnjaic and S.~D.~McDermott,
  {\it Severely Constraining Dark Matter Interpretations of the 21-cm Anomaly,}
  Phys.\ Rev.\ Lett.\  {\bf 121} (2018) no.1,  011102
  [arXiv:1803.02804 [hep-ph]].
  
\bibitem{Barkana:2018cct}
  R.~Barkana, N.~J.~Outmezguine, D.~Redigolo and T.~Volansky,
  {\it Strong constraints on light dark matter interpretation of the EDGES signal,}
  Phys.\ Rev.\ D {\bf 98} (2018) no.10,  103005
  [arXiv:1803.03091 [hep-ph]].
  
\bibitem{Fraser:2018acy}
  S.~Fraser {\it et al.},
  {\it The EDGES 21 cm Anomaly and Properties of Dark Matter,}
  Phys.\ Lett.\ B {\bf 785} (2018) 159
  [arXiv:1803.03245 [hep-ph]].
  
\bibitem{DAmico:2018sxd}
  G.~D'Amico, P.~Panci and A.~Strumia,
  {\it Bounds on Dark Matter annihilations from 21 cm data,}
  Phys.\ Rev.\ Lett.\  {\bf 121} (2018) no.1,  011103
  [arXiv:1803.03629 [astro-ph.CO]].
  
\bibitem{Hill:2018lfx}
  J.~C.~Hill and E.~J.~Baxter,
  {\it Can Early Dark Energy Explain EDGES?,''}
  JCAP {\bf 1808} (2018) no.08,  037
  [arXiv:1803.07555 [astro-ph.CO]].

\bibitem{Safarzadeh:2018hhg}
  M.~Safarzadeh, E.~Scannapieco and A.~Babul,
  {\it A limit on the warm dark matter particle mass from the redshifted 21 cm absorption line,}
  Astrophys.\ J.\  {\bf 859} (2018) no.2,  L18
  [arXiv:1803.08039 [astro-ph.CO]].
  
\bibitem{Hektor:2018qqw}
  A.~Hektor, G.~H\"{u}tsi, L.~Marzola, M.~Raidal, V.~Vaskonen and H.~Veermäe,
  {\it Constraining Primordial Black Holes with the EDGES 21-cm Absorption Signal,}
  Phys.\ Rev.\ D {\bf 98} (2018) no.2,  023503
  [arXiv:1803.09697 [astro-ph.CO]].
  
\bibitem{Slatyer:2018aqg}
  T.~R.~Slatyer and C.~L.~Wu,
 {\it Early-Universe constraints on dark matter-baryon scattering and their implications for a global 21 cm signal,}
  Phys.\ Rev.\ D {\bf 98} (2018) no.2,  023013
  [arXiv:1803.09734 [astro-ph.CO]].
  
\bibitem{Mitridate:2018iag}
  A.~Mitridate and A.~Podo,
  {\it Bounds on Dark Matter decay from 21 cm line,}
  JCAP {\bf 1805} (2018) no.05,  069
  [arXiv:1803.11169 [hep-ph]].
  
\bibitem{Munoz:2018jwq}
  J.~B.~Mu\~{n}oz, C.~Dvorkin and A.~Loeb,
  {\it 21-cm Fluctuations from Charged Dark Matter,}
  Phys.\ Rev.\ Lett.\  {\bf 121} (2018) no.12,  121301
  [arXiv:1804.01092 [astro-ph.CO]].
  
\bibitem{Witte:2018itc}
  S.~Witte, P.~Villanueva-Domingo, S.~Gariazzo, O.~Mena and S.~Palomares-Ruiz,
  {\it EDGES result versus CMB and low-redshift constraints on ionization histories,}
  Phys.\ Rev.\ D {\bf 97} (2018) no.10,  103533
  [arXiv:1804.03888 [astro-ph.CO]].
  
\bibitem{Li:2018kzs}
  C.~Li and Y.~F.~Cai,
  {\it Searching for the Dark Force with 21-cm Spectrum in Light of EDGES,}
  Phys.\ Lett.\ B {\bf 788} (2019) 70
  [arXiv:1804.04816 [astro-ph.CO]].
  
\bibitem{Jia:2018csj}
  L.~B.~Jia,
  {\it Dark photon portal dark matter with the 21-cm anomaly,}
  Eur.\ Phys.\ J.\ C {\bf 79} (2019) no.1,  80
  [arXiv:1804.07934 [hep-ph]].
  
\bibitem{Schneider:2018xba}
  A.~Schneider,
  {\it Constraining noncold dark matter models with the global 21-cm signal,}
  Phys.\ Rev.\ D {\bf 98} (2018) no.6,  063021
  [arXiv:1805.00021 [astro-ph.CO]].
  
\bibitem{Houston:2018vrf}
  N.~Houston, C.~Li, T.~Li, Q.~Yang and X.~Zhang,
  {\it Natural Explanation for 21 cm Absorption Signals via Axion-Induced Cooling,}
  Phys.\ Rev.\ Lett.\  {\bf 121} (2018) no.11,  111301
  [arXiv:1805.04426 [hep-ph]].
  
\bibitem{Wang:2018azy}
  Y.~Wang and G.~B.~Zhao,
  {\it Constraining the dark matter-vacuum energy interaction using the EDGES 21-cm absorption signal,}
  Astrophys.\ J.\  {\bf 869} (2018) no.1,  26
  [arXiv:1805.11210 [astro-ph.CO]].
  
\bibitem{Xiao:2018jyl}
  L.~Xiao, R.~An, L.~Zhang, B.~Yue, Y.~Xu and B.~Wang,
  {\it Can conformal and disformal couplings between dark sectors explain the EDGES 21-cm anomaly?,}
  Phys.\ Rev.\ D {\bf 99} (2019) no.2,  023528
  [arXiv:1807.05541 [astro-ph.CO]].
  
\bibitem{Kovetz:2018zan}
  E.~D.~Kovetz, V.~Poulin, V.~Gluscevic, K.~K.~Boddy, R.~Barkana and M.~Kamionkowski,
  {\it Tighter limits on dark matter explanations of the anomalous EDGES 21 cm signal,}
  Phys.\ Rev.\ D {\bf 98} (2018) no.10,  103529
  [arXiv:1807.11482 [astro-ph.CO]].
  
\bibitem{Jia:2018mkc}
  L.~B.~Jia, X.~J.~Deng and C.~F.~Liu,
  {\it Could the 21-cm absorption be explained by the dark matter suggested by$^8$ Be transitions?,}
  Eur.\ Phys.\ J.\ C {\bf 78} (2018) no.11,  956
  [arXiv:1809.00177 [hep-ph]].
  
\bibitem{Kovetz:2018zes}
  E.~D.~Kovetz, I.~Cholis and D.~E.~Kaplan,
  {\it Bounds on Ultra-Light Hidden-Photon Dark Matter from 21cm at Cosmic Dawn,}
  Phys.\ Rev.\ D {\bf 99} (2019) no.12,  123511
  [arXiv:1809.01139 [astro-ph.CO]].
  
\bibitem{Lopez-Honorez:2018ipk}
  L.~Lopez-Honorez, O.~Mena and P.~Villanueva-Domingo,
  {\it Dark matter microphysics and 21 cm observations,}
  Phys.\ Rev.\ D {\bf 99} (2019) no.2,  023522
  [arXiv:1811.02716 [astro-ph.CO]].
  
\bibitem{Nebrin:2018vqt}
  O.~Nebrin, R.~Ghara and G.~Mellema,
  {\it Fuzzy Dark Matter at Cosmic Dawn: New 21-cm Constraints,}
  JCAP {\bf 1904} (2019) no.04,  051
  [arXiv:1812.09760 [astro-ph.CO]].
  
\bibitem{Widmark:2019cut}
  A.~Widmark,
  {\it 21 cm cosmology and spin temperature reduction via spin-dependent dark matter interactions,}
  JCAP {\bf 1906} (2019) no.06,  014
  [arXiv:1902.09552 [astro-ph.CO]].
  
\bibitem{Li:2019loh}
  C.~Li, X.~Ren, M.~Khurshudyan and Y.~F.~Cai,
  {\it Implications of the possible 21-cm line excess at cosmic dawn on dynamics of interacting dark energy,}
  arXiv:1904.02458 [astro-ph.CO].

\bibitem{Aghanim:2015xee} 
  N.~Aghanim {\it et al.} [Planck Collaboration],
  {\it Planck 2015 results. XI. CMB power spectra, likelihoods, and robustness of parameters},
  Astron.\ Astrophys.\  {\bf 594}, A11 (2016)
  [arXiv:1507.02704 [astro-ph.CO]].
  
\bibitem{Lewis:2002ah} 
  A.~Lewis and S.~Bridle,
  {\it Cosmological parameters from CMB and other data: A Monte Carlo approach,}
  Phys.\ Rev.\ D {\bf 66}, 103511 (2002)
  [astro-ph/0205436].

\bibitem{gelmanrubin}
  A.~Gelman and D.~Rubin, 
  {\it Inference from iterative simulation using multiple sequences,} 
  Statistical Science \textbf{7}, 457 (1992).

\bibitem{Ade:2015xua}
  P.~A.~R.~Ade {\it et al.} [Planck Collaboration],
  {\it Planck 2015 results. XIII. Cosmological parameters,}
  Astron.\ Astrophys.\  {\bf 594}, A13 (2016)
  [arXiv:1502.01589 [astro-ph.CO]].

\bibitem{Riess:2019cxk}
  A.~G.~Riess, S.~Casertano, W.~Yuan, L.~M.~Macri and D.~Scolnic,
  {\it Large Magellanic Cloud Cepheid Standards Provide a 1\% Foundation for the Determination of the Hubble Constant and Stronger Evidence for Physics beyond $\Lambda$CDM,}
  Astrophys.\ J.\  {\bf 876} (2019) no.1,  85
  [arXiv:1903.07603 [astro-ph.CO]].
  
\bibitem{Vagnozzi:2019ezj}
  S.~Vagnozzi,
  {\it New physics in light of the $H_0$ tension: an alternative view,}
  arXiv:1907.07569 [astro-ph.CO].
  
\bibitem{Ade:2018sbj}
  P.~Ade {\it et al.} [Simons Observatory Collaboration],
  {\it The Simons Observatory: Science goals and forecasts,}
  JCAP {\bf 1902} (2019) 056
  [arXiv:1808.07445 [astro-ph.CO]].
  
\bibitem{Abitbol:2019nhf}
  M.~H.~Abitbol {\it et al.} [Simons Observatory Collaboration],
  {\it The Simons Observatory: Astro2020 Decadal Project Whitepaper,}
  Bull.\ Am.\ Astron.\ Soc.\  {\bf 51} (2019) 147
  [arXiv:1907.08284 [astro-ph.IM]].
  
\bibitem{Abazajian:2016yjj}
  K.~N.~Abazajian {\it et al.} [CMB-S4 Collaboration],
  {\it CMB-S4 Science Book, First Edition,}
  arXiv:1610.02743 [astro-ph.CO].
  
\bibitem{Sprenger:2018tdb}
  T.~Sprenger, M.~Archidiacono, T.~Brinckmann, S.~Clesse and J.~Lesgourgues,
  {\it Cosmology in the era of Euclid and the Square Kilometre Array,}
  JCAP {\bf 1902} (2019) 047
  [arXiv:1801.08331 [astro-ph.CO]].
  
\bibitem{Brinckmann:2018owf}
  T.~Brinckmann, D.~C.~Hooper, M.~Archidiacono, J.~Lesgourgues and T.~Sprenger,
  {\it The promising future of a robust cosmological neutrino mass measurement,}
  JCAP {\bf 1901} (2019) 059
  [arXiv:1808.05955 [astro-ph.CO]].

\end{thebibliography}
\end{document}